\documentclass{article}

\usepackage[english]{babel}
\usepackage[T1]{fontenc}
\usepackage{xspace}
\usepackage[utf8]{inputenc}
\usepackage{booktabs}
\usepackage{multirow}
\usepackage{color}
\usepackage{euscript}
\usepackage{amsfonts}
\usepackage{bm}
\usepackage{environ}
\usepackage{comment}
\usepackage{cancel}

\usepackage[letterpaper,top=2cm,bottom=2cm,left=3cm,right=3cm,marginparwidth=1.75cm]{geometry}

\usepackage{amsmath}
\usepackage{graphicx}
\usepackage{caption}
\usepackage[colorlinks=true, allcolors=blue]{hyperref}
\usepackage{subfigure}

\def\bmr{{\bm r}}
\def\bmp{{\bm p}}
\def\bmk{{\bm k}}
\def\bmP{{\bm P}}
\def\bmR{{\bm R}}
\def\bmt{{\bm t}}

\title{ToMCCA-3: A realistic 3-body coalescence model}
\author{Maximilian Mahlein$^{1}$, Bhawani Singh$^{5}$, Michele Viviani$^{2}$,\\ Francesca Bellini$^{3}$, Laura Fabbietti$^{1}$, Alejandro Kievsky$^{2}$,\\Laura Elisa Marcucci $^{4,2}$
\vspace{0.2cm}
\\$^1$Technical University of Munich, TUM School of Natural Sciences, Physics Department,\\ James-Franck-Stra{\ss}e 1, 85748 Garching b. M\"{u}nchen, Germany
 \\$^2$ INFN Pisa, Largo B. Pontecorvo 3, I-56127 Pisa, Italy
 \\$^3$ Dipartimento di Fisica e Astronomia A. Righi, Universit\`a di Bologna and INFN\\ Sezione di Bologna, Via Irnerio 46, 40126 Bologna, Italy
 \\$^4$ Dipartimento di Fisica. Universit\`a di Pisa, Largo B. Pontecorvo 3, I-56127 Pisa, Italy
 \\$^5$ Thomas Jefferson National Accelerator Facility, Newport News, VA 23606, USA
 }

\begin{document}
\maketitle

\begin{abstract}
The formation of light (anti)nuclei in high-energy collisions provides valuable insights into the underlying dynamics of the strong interaction and the structure of the particle-emitting source. Understanding this process is crucial not only for nuclear physics but also for astrophysical studies, where the production of rare antinuclei could serve as a probe for new physics. This work presents a three-body coalescence model based on the Wigner function formalism, offering a refined description of light-(anti)nuclei production in ultra relativistic proton-proton collisions. By incorporating realistic two- and three-body nuclear interaction potentials constrained by modern scattering and femtoscopic correlation data, our approach improves on traditional coalescence models. The framework is validated using the ToMCCA event generator applied to proton-proton collisions at $\sqrt{s}=13$~TeV to predict the momentum spectra of light (anti)(hyper)nuclei with mass number $A=3$, which are then compared with the experimental data from ALICE. Our results demonstrate the sensitivity of light nuclei yields to the choice of nuclear wave functions, emphasizing the importance of an accurate description of the coalescence process. This model lays the foundation for the extension of coalescence studies of $A=3$ light nuclei to a wider range of collision systems and energies.

\end{abstract}

\newcommand{\sk}[1]{\textcolor{cyan}{SK: #1}}

\def\tri{{{}^3{\rm H}}}
\def\hyp{{{}^3_{\Lambda}{\rm H}}}
\def\hel{{{}^3{\rm He}}}
\def\het{{{}^3{\rm He}}}
\def\heq{{{}^4{\rm He}}}
\def\lis{{{}^7{\rm Li}}}
\def\bes{{{}^7{\rm Be}}}
\def\beo{{{}^8{\rm Be}}}
\def\boo{{{}^8{\rm B}}}
\def\a{\alpha}
\def\b{\beta}
\def\bmr{{\bm r}}
\def\bmx{{\bm x}}
\def\bmy{{\bm y}}
\def\bmz{{\bm z}}
\def\bmp{{\bm p}}
\def\bmq{{\bm q}}
\def\bme{{\bm e}}
\def\bmj{{\bm j}}
\def\bmP{{\bm P}}
\def\bmk{{\bm k}}
\def\bmK{{\bm K}}
\def\bmt{{\bm t}}
\def\bmZ{{\bm Z}}
\def\htm{{\hbar^2 \over M}}
\def\htm{{\hbar}}
\def\np{\phantom{0}}
\def\jac{\xi}
\newcommand{\jacb}{{\bm \xi}}
\def\hypfi{\varphi}

\newcommand{\MeVc}{\ensuremath{\mathrm{MeV}\kern-0.05em/\kern-0.02em \textit{c}}~}
\newcommand{\GeVc}{\ensuremath{\mathrm{GeV}\kern-0.05em/\kern-0.02em \textit{c}}~}
\newcommand{\GeVcSq}{\ensuremath{\mathrm{GeV}\kern-0.05em/\kern-0.02em \textit{c}^2}~}
\newcommand{\MeVcSq}{\ensuremath{\mathrm{MeV}\kern-0.05em/\kern-0.02em \textit{c}^2}~}
\newcommand{\fermi}{\ensuremath{\mathrm{fm}}~}
\newcommand{\Led}{Lednick\'y\xspace}
\newcommand{\HALQCD}{HAL QCD\xspace}
\newcommand{\ChargedTrackMultiplicity}{\ensuremath{\num{35}}\,}
\newcommand{\ResolutionEffect}{8}

\newcommand{\ie}{\textit{i.e.}}
\newcommand{\eg}{\textit{e.g.}}
\newcommand{\cf}{\textit{cf.}}

\newcommand{\MeV}{\ensuremath{\mathrm{MeV}}}
\newcommand{\fm}{\ensuremath{\mathrm{fm}}}

\newcommand{\MN}{M_N}
\newcommand{\Md}{M_d}
\newcommand{\Mpi}{M_\pi}

\newcommand{\dd}{\mathrm{d}}
\newcommand{\ii}{\ensuremath{i}}
\newcommand{\ee}{\ensuremath{e}}

\newcommand{\vD}{\boldsymbol{D}}
\newcommand{\hc}{\mathrm{h.c.}}

\newcommand{\vk}{\mathbf{k}}
\newcommand{\vp}{\mathbf{p}}
\newcommand{\vq}{\mathbf{q}}

\newcommand{\vecr}{\mathbf{r}}
\newcommand{\vecx}{\mathbf{x}}
\newcommand{\vecy}{\mathbf{y}}
\newcommand{\vecz}{\mathbf{z}}
\newcommand{\vecn}{\mathbf{n}}
\newcommand{\vecm}{\mathbf{m}}
\newcommand{\vecp}{\mathbf{p}}
\newcommand{\veck}{\mathbf{k}}
\newcommand{\vecq}{\mathbf{q}}
\newcommand{\vece}{\mathbf{e}}
\newcommand{\veca}{\mathbf{a}}
\newcommand{\vecb}{\mathbf{b}}
\newcommand{\vecu}{\mathbf{u}}
\newcommand{\vecv}{\mathbf{v}}

\newcommand{\couple}[3]{\left({#1}{#2}\right)\!{#3}}

\NewEnviron{subalign}[1][]{%
\begin{subequations}\begin{align}
  \BODY
\end{align}\label{#1}\end{subequations}
}

\NewEnviron{spliteq}{%
\begin{equation}\begin{split}
  \BODY
\end{split}\end{equation}
}

\newcommand{\abs}[1]{\left|#1\right|}

\newcommand{\TopRule}{\toprule[1.25pt]}
\newcommand{\BottomRule}{\TopRule}
\newcommand{\MidRule}{\midrule[0.5pt]}
\newcommand{\MidDoubleRule}{\hline\hline}

\renewcommand{\vec}[1]{\mathbf{#1}}

\newcommand{\mmt}{\ensuremath{\langle m_\mathrm{T}\rangle}\xspace}
\newcommand{\AV}{\ensuremath{\text{Argonne }v_{18}}\xspace}
\newcommand{\dNetaOFive}{\ensuremath{\langle\diff N_\mathrm{ch}/\diff\eta\rangle_{|\eta|<0.5}}\xspace}
\newcommand{\heThree}          {\ensuremath{{}^3}\text{He}\xspace}
\newcommand{\hThree}          {\ensuremath{{}^3}\text{H}\xspace}
\newcommand{\lhThree}          {\ensuremath{{}^3_\Lambda}\text{H}\xspace}
\newcommand{\pt}{\ensuremath{p_\mathrm{T}}\xspace}
\newcommand{\diff}{\ensuremath{\mathrm{d}}\xspace}

\newcommand{\pL}{N\Lambda}
\newcommand{\ppL}{NN\Lambda}
\newcommand{\bmS}{\bm{S}} 
\section{Introduction}
One of the fundamental open questions in nuclear and particle physics, as well as astroparticle physics, is understanding the formation mechanisms of light nuclei and antinuclei—such as deuterons and helium—in high-energy interactions. In recent years, research on antimatter nuclei heavier than antiprotons has seen significant advancements on both experimental and theoretical fronts.

This question has received particular attention at high energy hadron colliders such as the Large Hadron Collider (LHC) and the Relativistic Heavy Ion Collider (RHIC). Due to the large integrated luminosities delivered over the past decade, multi-differential measurements of light nuclei and antinuclei production could be performed with unprecedented precision. New techniques, such as femtoscopy~\cite{Lisa:2005dd,Fabbietti:2020bfg} and correlation studies~\cite{ALICECollaboration2024Sep,ALICECollaboration2025Apr}, have provided deeper insights into the production mechanisms of these bound states. 

Light (anti-)nuclei yields and particle ratios measured as functions of the final-state charged-particle multiplicity pseudorapidity density (or \textit{multiplicity}) have shown the dependence of nucleus production on the size of the particle emitting source \cite{ALICE:2019dgz, ALICE:2019bnp, ALICE:2021ovi, ALICE:2021mfm, ALICE:2022veq}. 
Different particle species, among which nucleons and strange baryons, are observed to be emitted from a common particle source in proton-proton collisions at center of mass energies of $\sqrt{s} = 13$ TeV~\cite{ALICE:2020ibs, ALICE:2023sjd}, whose radius decreases with transverse momentum\footnote{More precisely, the one dimensional radius assuming a Gaussian source profile decreases with the average particle pair transverse momentum, $k_T$.}. In addition, pion kaon and proton radii are observed to increase with multiplicity from pp to p--Pb (\textit{small systems}) to heavy-ion collisions (\textit{large systems})~\cite{ALICE:2012aai, ALICE:2015hvw}. 
Moreover, proton-proton (p--p), proton-deuteron (p--d), Lambda-deuteron ($\Lambda$--d) and three particle correlations (p--p--p and p--p--$\Lambda$) studied with femtoscopy have allowed access to the strong interaction between baryons, shedding light on the three-body forces acting in systems of two and three nucleons \cite{ThreeBodyPPP}.

Among the light nuclei and antinuclei, deuterons have been measured with the highest precision, as a function of multiple variables such as multiplicity, transverse momentum, and rapidity. Models attempting to describe their formation—including statistical hadronization \cite{Andronic:2010qu, Andronic:2017pug, Vovchenko:2018fiy} and coalescence \cite{Butler:1963pp, Csernai:1986qf, Nagle:1996vp, Scheibl:1998tk, Blum:2019suo}, a nuclear fusion-like process in high-energy collisions -- have been largely successful, despite relying on different underlying principles. 
In statistical hadronization models (SHM), light nuclei are produced from a hadron gas in thermal equilibrium, according to the laws of quantum statistics, and with an appropriate treatment of the quantum number conservation laws, depending on the size of the system. SHM-predicted abundances depend on the particle mass and spin, as well as the temperature and volume of the system. 
The coalescence approach describes how individual (anti)nucleons merge into (anti)nuclei based on their proximity in momentum and spatial coordinates.
However, discrepancies between the measured and predicted observables remain, particularly in the production of heavier systems such as helium, triton, and hypertriton (a bound state of a neutron, a proton, and a $\Lambda$ baryon) \cite{ALICE:2022veq, ALargeIonColliderExperiment:2021puh}. The latter confirms to be a key observable for discriminating between thermal and coalescence production~\cite{Steinheimer:2012tb, Bellini:2018epz, Bellini2021} due to its broad size ($\sim10$~fm) and small energy that binds the $\Lambda$ to the deuteron core (world average $B_{\Lambda}=105\pm26 $~keV, \cite{HyperNucleiDatabase}). The first hypertriton measurement in p--Pb collisions, supports the coalescence model and disfavors the hypothesis of thermal production~\cite{ALargeIonColliderExperiment:2021puh}.

While widely used, coalescence models often involve free parameters, such as the coalescence momentum $p_0$, whose values are not derived from first principles and vary depending on the production environment. Traditional models assume simple momentum-based coalescence, while more advanced quantum-mechanical approaches incorporate nucleus wave functions and the size of the emission source, according to a Wigner function-based formulation \cite{Scheibl:1998tk, Blum:2019suo, Bellini2021}. Such an approach also finds application in recent developments based on Monte Carlo methods that allow us to apply coalescence as an afterburner to particle production~\cite{Kachelrie2020,Mahlein2023}.
One of the most intriguing applications of the research on (anti)nuclei formation mechanisms is the indirect detection of dark matter (DM) through cosmic antinuclei, such as antideuterons and antihelium. 
Unlike standard cosmic ray antimatter (antiprotons and positrons), these antinuclei have a significantly suppressed background from astrophysical sources, making them a promising signature of DM annihilation \cite{Donato:1999gy, Ibarra:2012cc, He3Cosmicrays, Carlson:2014ssa, Korsmeier:2017xzj,  Blum:2017qnn}. However, their production mechanisms remain uncertain, and measurements from particle accelerators become a key input.
The tentative measurements of $^{3,4}\overline{\mathrm{He}}$ antinuclei in cosmic rays by the AMS-02 Collaboration have sparked several studies in order to interpret these findings. Although the experimental results have not been published in a peer-reviewed journal yet, the confirmation of even one such antinucleus would challenge the current understanding of cosmology~\cite{vonDoetinchem2020,ALICEHeThreeCR}. 

In this paper, the coalescence model ToMCCA~\cite{Mahlein2024} is extended to the $A=3$ case.
ToMCCA is a lightweight event-by-event coalescence Monte-Carlo generator designed to study light-(anti)nuclei formation without the overhead of a full event generator and with easy customization. It includes parameterizations of the multiplicity dependence of proton transverse-momentum spectra and of the baryon-emitting source size tuned to (anti)deuteron measurements by ALICE in pp collisions at $\sqrt{s}=5-13\,\mathrm{TeV}$, as well as event-multiplicity distributions taken from EPOS. ToMCCA reproduces all the deuteron \pt distributions by ALICE within $\sim5\%$.

The predictions include not only the nuclei \heThree and \hThree, but also the hypernucleus \lhThree. The Wigner function formalism has been implemented for $A=3$ (anti)nuclei using realistic wave functions based on the pair-correlated hyperspherical harmonics (PHH) method~\cite{Kievsky:2008es}. For 2-body nucleon-nucleon (NN) interactions, the \AV potential (AV18)~\cite{Wiringa:1994wb} is used, together with the  Urbana IX (UIX) 3-body interaction~\cite{Pudliner:1995wk}. Furthermore, the Minnesota potential~\cite{MinnesotaPotential} is tested as a simple 2-body potential which is able to reproduce the binding energies of \heThree and \hThree. For \lhThree, a simplified approach is considered: the Congleton~\cite{Congleton_1992} approach is a simplified model of the \lhThree, with a deuteron at its core, modeled using the \AV wave function of the deuteron, and a shallowly bound $\Lambda$. The parameters of this model were tuned to reproduce modern form factor calculations~\cite{Bellini2021,Hildenbrand2019Sep}, and it uses a $\Lambda$ binding energy of $B_\Lambda=128$ keV, well in line with the world average.


\section{Coalescence approach for A = 3 nuclei}
The momentum spectra of light (anti)(hyper)nuclei\footnote{The formation mechanism is assumed to be the same for nuclei and their antimatter counterpart, as well as for hypernuclei. In the following, the (anti) and (hyper) prefixes will be omitted.} with mass number $A=3$, such as \hThree, \heThree, and \lhThree, can be computed using the coalescence formalism. In this framework, light nuclei are assumed to form in high-energy hadronic collisions through the coalescence of nucleons and hyperons. A key assumption is that interactions between coalescing hadrons and non-participating hadrons are subdominant. We develop a coalescence model for $A=3$ light nuclei based on the Wigner function formalism~\cite{Kachelrie2020,Mahlein2023}. We compute Lorentz-invariant momentum spectra of \hThree, \heThree, and \lhThree using the three-particle density matrix formalism as follows:\footnote{Note that in these calculations, we use bold and italic fonts for three- and four-vectors.}
\begin{multline}
\gamma \frac{\mathrm{d} N_{\mathrm{A}}}{\mathrm{~d}^3 P}=  \frac{S_{\mathrm{A}}}{(2 \pi)^4} \int \mathrm{~d}^4 x_1 \int \mathrm{~d}^4 x_2,\int \mathrm{~d}^4 x_3 \int \mathrm{~d}^4 x_1^{\prime} \int \mathrm{d}^4 x_2^{\prime}\int \mathrm{d}^4 x_3^{\prime} \\
 \times \Psi^*\left(x_1^{\prime}, x_2^{\prime}, x_3^{\prime}, \right) \Psi\left(x_1, x_2,x_3\right) \rho_{1,2,3}\left(x_1, x_2, x_3 ; x_1^{\prime}, x_2^{\prime}, x_3^{\prime}\right),
 \label{eq:Eq1}
\end{multline}
where $\Psi\left(x_1, x_2,x_3\right) $ is the three-particle bound state Bethe-Salpeter amplitude and $\rho_{1,2,3}$ is the reduced three-particle density matrix, while $S_\mathrm{A}$ accounts for the spin and isospin statistics for all three cases\footnote{$S_A=1/12$ for \heThree and \hThree and 1/8 for \lhThree. The latter is due to the non-existent isospin of the $\Lambda$}. The density matrix for three particle systems is assumed to be factored into single particle densities~\cite{Bellini2021}, $\rho_{1,2,3}\left(x_1, x_2, x_3; x_1^{\prime}, x_2^{\prime}, x_3^{\prime}\right)= \rho_{1}\left(x_1; x_1^{\prime}\right)\times \rho_{1}\left(x_2; x_2^{\prime}\right)\times \rho_{1}\left(x_3; x_3^{\prime}\right)$, and the single-particle density is written in terms of the single particle Wigner function, $f_1^W$. One can make the assumption

\begin{eqnarray}
    \rho_1\left(x, x^{\prime}\right)=\int \frac{\mathrm{d}^4 p}{(2 \pi)^4} e^{i p\cdot\left(x^{\prime}-x\right)} f_1^W\left(p, \frac{x+x^{\prime}}{2}\right) .
\end{eqnarray}
Furthermore, the three-particle normalized Wigner density function, $W_\mathrm{A}$, can be defined as a product of single-particle Winger density functions
\begin{eqnarray}
    W_\mathrm{A}\left( p_1, p_2, p_3,\frac{x_1 +x_1^\prime}{2},\frac{x_2+x_2^\prime}{2},\frac{x_3 +x_3^\prime}{2}\right) = \prod_{i =1}^{3}
    f_1^W\left(p_i, \frac{x_i+x_i^{\prime}}{2}\right)\,.
\end{eqnarray}

At this stage, making both the low-energy approximation and the equal-time approximation (see App.~\ref{sec:4dto3dreduction}), Eq.~(\ref{eq:Eq1}) takes the form of a yield equation for $A=3$ cluster in three-dimensional space. The yield in three-dimensional space is written as 
\begin{multline}
 \frac{\mathrm{d} N_{\mathrm{A}}}{\mathrm{~d}^3 P}=  \frac{S_{\mathrm{A}}}{(2 \pi)^{12}}\int \mathrm{d}^3 \bm{p}_1 \int \mathrm{d}^3 \bm{p}_2 \int \mathrm{d}^3 \bm{p}_3 \int \mathrm{~d}^3 \bm{x}_1 \int \mathrm{~d}^3 \bm{x}_2\,\int \mathrm{~d}^3 \bm{x}_3 \int \mathrm{~d}^3 \bm{x}_1^{\prime} \int \mathrm{d}^3 \bm{x}_2^{\prime}\int \mathrm{d}^3 \bm{x}_3^{\prime} \\ \times \Psi^*_{NR}\left(\bm{x}_1^{\prime}, \bm{x}_2^{\prime}, \bm{x}_3^{\prime}, \right) \Psi_{NR}\left(\bm{x}_1, \bm{x}_2,\bm{x}_3\right) \,e^{-i \bm{p}_1\cdot\left(\bm{x}_1 -\bm{x}_1^\prime\right)-i \bm{p}_2\cdot\left(\bm{x}_2 -\bm{x}_2^\prime\right)-i \bm{p}_3\cdot\left(\bm{x}_3 -\bm{x}_3^\prime\right)}\,\\\times W_\mathrm{A}\left( \bm{p}_1, \bm{p}_2, \bm{p}_3,\frac{\bm{x}_1 +\bm{x}_1^\prime}{2},\frac{\bm{x}_2+\bm{x}_2^\prime}{2},\frac{\bm{x}_3 +\bm{x}_3^\prime}{2}\right)\,,
 \label{eq:yieldEq2}
\end{multline}
where $\Psi_{NR}$ is the non-relativistic wave function of the nucleus under study.
In general, it can be factorized into the plane wave for center-of-mass motion with total momentum $\bm{P}_\mathrm{A}$ and an internal wave function that depends only on two relative coordinates $\bm \xi_1$, $\bm \xi_{2}$ describing the motion of internal nucleons or $\Lambda$ relative to the center of mass,
\begin{eqnarray}
    \Psi_\mathrm{NR}\left(\bm{x}_1, \bm{x}_2, \bm{x}_3\right) = \left(2\pi\right)^{-3/2} e^{i \bm{P}_\mathrm{A}\cdot \bm{R}}\varphi_\mathrm{NR}(\bm{\xi}_1,\bm{\xi}_2)
    \label{eq:Wavefunction_A}\,,
\end{eqnarray}
The relative Jacobi coordinates, $\bm{\xi}_1$, $\bm{\xi}_{2}$  are defined in terms of the constituents absolute coordinates $\bm{x}_1,\bm{x}_2$, and $\bm{x}_3$, as well as the masses of the three particles $m_1,m_2$, and $m_3$. Moreover, $\bm R$ represents the center-of-mass coordinate. The internal cluster wave function is denoted by $\varphi_{\mathrm{NR}}(\bm \xi_1, \bm \xi_{2})$. Considerations regarding the choice of Jacobi coordinates for $A=3$ systems are discussed in subsequent parts of this paper.\\

Let us first consider a system consisting of two identical particles with masses $m_1=m_2=M$ and a third particle with a different mass $m_3$. We define the mass ratio as $\kappa=\frac{m_3}{M}$ so that the cases of \heThree and \hThree can be simply recovered by setting $\kappa=1$ and the case of \lhThree by setting $\kappa=\frac{m_\Lambda}{M}\approx1.188$ in the final expressions.
The Jacobi vectors are
\begin{equation}
    \bmR={\bmx _1+\bmx _2+\kappa\bmx_3\over 2+\kappa}\,,\,\,
    \bm{\xi}_2= \bmx _2 -\bmx _1\,,\,\,\text{and}\,\,
     \bm{\xi}_1=\sqrt{4\kappa\over 2+\kappa} \left( \bmx _3 - {\bmx _1+\bmx_2\over 2}\right) \ ,
\end{equation}

\begin{equation}
    \begin{pmatrix}
 \bm{\xi}_1 \\ \bm{\xi}_2 \\ \bmR 
\end{pmatrix}=J \times \begin{pmatrix}
    \bmx_{1} \\ \bmx_{2} \\ \bmx_{3} 
\end{pmatrix}\ , \,\,
   \begin{pmatrix}
  \bmx_{1} \\ \bmx_{2} \\ \bmx_{3} 
\end{pmatrix}=J^{-1} \times \begin{pmatrix}
 \bm{\xi}_1 \\ \bm{\xi}_2 \\ \bmR   
\end{pmatrix},\text{and}\, J =\left(
\begin{array}{ccc}
 -{1\over 2}\sqrt{4\kappa\over 2+\kappa } &  -{1\over 2}\sqrt{4\kappa\over 2+\kappa } &  \sqrt{4\kappa\over 2+\kappa } \\
 -1 & 1 & 0 \\
  {1\over 2+\kappa}  & {1\over 2+\kappa} & {\kappa\over 2+\kappa} \\
\end{array}
\right)\,,
\end{equation}
with an absolute value of the determinant as $|J| = \sqrt{4\kappa \over 2+\kappa}$. In terms of the Jacobi coordinates,
\begin{equation}
    \bmx_{13}=-\sqrt{2+\kappa\over 4\kappa} \bm{\xi}_1-{1\over 2}\bm{\xi}_2\ ,\qquad \bmx_{32}=\sqrt{2+\kappa\over 4\kappa} \bm{\xi}_1-{1\over 2}\bm{\xi}_2\ .
\end{equation}
The associated Jacobi momenta are
\begin{equation}
    \bmP=\bmp_1+\bmp_2+\bmp_3\ ,\qquad
    \bm{k}_2= {1\over 2}\Bigl[ \bmp_2 -\bmp_1\Bigr]\ ,\,
     \bm{k}_1=\sqrt{\kappa\over 2+\kappa } \left[ {\bmp_3\over \kappa} -  {\bmp_1+\bmp_2\over 2}\right] \ .
\end{equation}
The positions and momenta in terms of the Jacobi coordinates and momenta are
\begin{eqnarray}
\bmx_1&= \bmR-{1\over2}\sqrt{\kappa\over 2+\kappa} \bm{\xi}_1 +{\bm{\xi}_2\over 2}\ ,
  & \qquad\bmp_1 = {1\over 2+\kappa} \bmP -{1\over 2}\sqrt{4\kappa\over 2+\kappa}\bmk_1+\bmk_2 \\
\bmx_2&= \bmR-{1\over2}\sqrt{\kappa\over 2+\kappa} \bm{\xi}_1 -{\bm{\xi}_2\over 2}\ ,
  & \qquad \bmp_2 = {1\over 2+\kappa} \bmP -{1\over 2}\sqrt{4\kappa\over 2+\kappa}\bmk_1-\bmk_2 \\
\bmx_3&= \bmR+{1\over \kappa}\sqrt{\kappa\over 2+\kappa} \bm{\xi}_1 \ ,
  & \qquad \bmp_3 = {\kappa\over 2+\kappa} \bmP +\sqrt{4\kappa\over 2+\kappa}\bmk_1\ .
\end{eqnarray}
From the definitions of the Jacobi coordinates, the following relation immediately follows.
\begin{equation}
    \bmx_1\cdot\bmp_1+ \bmx_2\cdot\bmp_2+ \bmx_3\cdot\bmp_3=
    \bmR\cdot\bmP+\bm{\xi}_1\cdot\bmk_1+\bm{\xi}_2\cdot\bmk_2\ .
\end{equation}

The complete form of the Wigner function $W_\mathrm{A}$ is theoretically unknown~\cite{Kachelrie2020}. However, by assuming a transition from a fully quantum mechanical treatment to a semi-classical picture and a factorization of the spatial and momentum coordinates
the Wigner function $W_\mathrm{A}$ can be expressed as the product of two normalized functions: the spatial distribution $H_\mathrm{A}$ and the momentum distribution $G_\mathrm{A}$ of the coalescing hadrons
\begin{eqnarray}
W_\mathrm{A}\left( \bm{p}_1, \bm{p}_2, \bm{p}_3,\bm{x}_1,\bm{x}_1^\prime,\bm{x}_2,\bm{x}_2^\prime,\bm{x}_3,\bm{x}_3^\prime\right)=H_\mathrm{A}\left(\bm{x}_1,\bm{x}_1^\prime,\bm{x}_2,\bm{x}_2^\prime,\bm{x}_3,\bm{x}_3^\prime\right)G_\mathrm{A}\left( \bm{p}_1, \bm{p}_2, \bm{p}_3\right)\,,
\end{eqnarray}
where  the function$H_\mathrm{A} $ is assumed to be the product of Gaussian distributions $H_\mathrm{A}  = \prod_{i=1}^\mathrm{A} h(x_i)$ and $h(x_i)$ is a normalized Gaussian distribution, 
\begin{eqnarray}
    h(\boldsymbol{x})=\left(2 \pi \sigma^2\right)^{-3 / 2} \exp \left\{-\frac{x^2}{2 \sigma^2}\right\}\,.
\end{eqnarray}

The width parameter $\sigma$ is related to the particle emission profile in the ultra-high-energy hadronic collisions and can be obtained from two-particle femtoscopic source studies in proton-proton collisions~\cite{ALICE:2023sjd, Acharya2020Dec, pion-protonPaper}. The exact form of the function $G_\mathrm{A}$ is still not known theoretically; however, since it is a distribution related to the momentum spectra of the individual nucleons, the values of this function can be incorporated by employing an event generators~\cite{Kachelrie2020,Mahlein2023}. \\
\subsection{Generalized 3-body coalescence approach}
In the following, we will derive the calculation of the coalescence probability for an arbitrary $A=3$ (anti)(hyper)nucleus. This defines a generalized approach for 2 equal (nucleons) and one different (hyperon) particle. Under the assumption that the third particle is equal to the other two, the cases of \heThree and \hThree are recovered.
The Wigner density matrix for the system of $\mathrm{np}\Lambda$ appearing in Eq.~(\ref{eq:yeildEq2}) can be written as
\begin{eqnarray}
    W_{3}\left( \bm p_1, \bm p_2, \bm p_3,\frac{\bmx_1 +\bmx_1^\prime}{2},\frac{\bmx_2+\bmx_2^\prime}{2},\frac{\bmx_3 +\bmx_3^\prime}{2}\right) &= &{1\over (2\pi\sigma^2)^3} \exp\left(-{ (\bmx_1+\bmx_1')^2+(\bmx_2+\bmx_2')^2\over 8\sigma^2}\right)\\
    &&\times{1\over (2\pi\sigma'^2)^{3/2}}\exp\left(-{ (\bmx_3+\bmx_3')^2\over 8\sigma'^2}\right) G_{3}(\bmp_1,\bm p_2,\bmp_3)\ ,
    \label{eq:WignerDensityHe}
\end{eqnarray}
where $ \sigma $ and $ \sigma' $ represent the femtoscopic source sizes for  N--N  pairs and e.g. N--$\Lambda $ pairs, respectively, and are assumed to be independent of each other. The values of $ \sigma $ and $ \sigma' $ are constrained by recent femtoscopic source size measurements in pp collisions~\cite{ALICE:2023sjd, Acharya2020Dec, pion-protonPaper} and can be obtained from event generators, which are tuned to reproduce the data; more details are discussed in the subsequent sections.
Defining 
$ \tau = (\sigma / \sigma')^2 $ yields the following expressions for the Gaussian forms,
\begin{equation}
  \exp\left(-{(\bmx_1+\bmx_1')^2+(\bm x_2+\bm x_2')^2\over 8\sigma^2}-{(\bm x_3+\bm x_3')^2\over 8\sigma^{\prime 2}}\right)
  = \exp\left(-{(\bm x_1+\bm x_1')^2+(\bm x_2+\bm x_2')^2+\tau\, (\bm x_3+\bm x_3')^2\over 8\sigma^2}\right)   \ .\label{eq:expo}
\end{equation}
The exponent of Eq.~(\ref{eq:expo}) can be expressed in terms of  $\kappa$, $\tau$, and dummy variables $\bmt$ and $\alpha$
\begin{equation}
    (\bmx_1+\bmx_1')^2+(\bmx_2+\bmx_2')^2+\tau\; (\bmx_3+\bmx_3')^3= (2+\tau) \Bigl[(\bmR+\bmR')^2 + 2 (\bmR+\bmR')\cdot\bmt\Bigr] +{\alpha\over2} (\bm{\xi}_1+\bm{\xi}_1')^2+{1\over2}(\bm{\xi}_2+\bm{\xi}_2')^2\ ,
\end{equation}
where $\bmt$ and $\alpha$ are defined as
\begin{equation}
    \bmt={\tau-\kappa\over \kappa(2+\tau)}\sqrt{\kappa\over 2+\kappa} (\bm{\xi}_1+\bm{\xi}_1')\ , \qquad
    \alpha={\kappa^2+2\tau\over \kappa(2+\kappa)} \ .
    \label{eq:DeftandAlpha}
\end{equation}
As discussed before, the wave function for the $A=3$ nucleus can be written in terms of the center-of-mass motion and an internal wave function as 
\begin{equation}
    \Psi_{NR}\left(\bmx_1,\bmx_2,\bmx_3\right) = \left(2\pi\right)^{-3/2} e^{i\bmP_{3}\cdot \bmR}\varphi_{{3}}(\bm{\xi}_1,\bm{\xi}_2)\ ,
    \label{eq:WavefunctionHyp}
\end{equation}
where $\varphi_{3}$ is the internal part of the $A=3$ wave function, describing the dynamics of the p, n, and the third baryon and $\bm \xi_1$ and $\bm \xi_2$ are the Jacobi coordinates defined above.
In Eq.~(\ref{eq:yieldEq2}) changing the integration variables from the particle positions and momenta to the center-of-mass and Jacobi coordinates, one obtains
\begin{eqnarray}
    \frac{\mathrm{d} N_3}{\mathrm{d}^3 P}&=&S_{3} {|J^{-1}|^3\over (2\pi)^{15}}\int \mathrm{d}^3 \bm{R} \,\mathrm{d}^3 \bm{\xi}_1 \,\mathrm{d}^3 \bm{\xi}_2\int \mathrm{d}^3 \bm{R}^{\prime} \,\mathrm{d}^3 \,\bm{\xi}_1^{\prime} \mathrm{d}^3 \bm{\xi}_2^{\prime}\, 
    \exp\left({i\bm{P}_{3}\cdot (\bmR-\bmR')}\varphi_{3}^*(\bm{\xi}_1',\bm{\xi}_2')\varphi_{3}(\bm{\xi}_1,\bm{\xi}_2)\right)\nonumber \\ 
    && \times\int \mathrm{d}^3 \bm{P} \,\mathrm{d}^3 \bm{k}_1 \,\mathrm{d}^3 \bm{k}_2\, 
   \exp\left({-i \bm{P}\cdot\left(\bmR-\bmR'\right)-i \bmk_1 \cdot\left(\bm{\xi}_1 -\bm{\xi}_1'\right)-i \bmk_2 \cdot\left(\bm{\xi}_2 -\bm{\xi}_2^\prime\right)}\right) \nonumber \\
   &&\times {\tau^{3\over2}\over (2\pi\sigma^2)^{9\over2}}
   e^{-{2+\tau\over 8\sigma^2}\bigl((\bmR+\bmR')^2+2(\bmR+\bmR')\cdot\bmt\bigr)}\;
   \exp\left({-{1\over 16\sigma^2}\bigl(\alpha(\bm{\xi}_1+\bm{\xi}_1')^2+(\bm{\xi}_2+\bm{\xi}_2')^2
   \bigr)}\right) \nonumber \\
   &&\times G_{3}(\bm p_1,\bm p_2,\bm p_3) \ ,
    \label{eq:main2}
\end{eqnarray}
where $|J^{-1}|=\sqrt{(2+\kappa)/4\kappa}$. Furthermore, by substituting $\bmR$ and $\bmR'$ with the variables $\bmZ=\bmR+\bmR'$ and $\bmZ'=\bmR-\bmR'$, the integration over $\bmZ'$ results in a factor $(2\pi)^3\delta(\bmP-\bmP_{3})$, while the integral over $\bmZ$ results in 
\begin{equation}
    \int \mathrm{d}^3\bm{Z}\; \exp\left(-{2+\tau\over 8 \sigma^2}[Z^{2}+2\bmt\cdot\bmZ]\right)
    = \int \mathrm{d}^3\bm{Z}\; \exp\left(-{2+\tau\over 8 \sigma^2}[(\bmZ+\bmt)^2-t^2]\right)=
    \left({8\pi\sigma^2\over 2+\tau}\right)^{3\over2}  \exp\left({2+\tau\over 8 \sigma^2} t^2\right)\ .
    \label{eq:dummyInt}
\end{equation}
Using these results,  Eq.~(\ref{eq:main2}) becomes
\begin{eqnarray}
    \frac{\mathrm{d} N_3}{\mathrm{d}^3 P}&=&{S_{3}\over 8} {|J^{-1}|^3\over (2\pi)^{12}}\int \mathrm{d}^3 \bm{\xi}_1 \,\mathrm{d}^3 \bm{\xi}_2\int \mathrm{d}^3 \,\bm{\xi}_1^{\prime} \mathrm{d}^3 \bm{\xi}_2^{\prime}\, 
   \varphi_{3}^*(\bm{\xi}_1',\bm{\xi}_2')\varphi_{3}(\bm{\xi}_1,\bm{\xi}_2)\nonumber \\ 
    && \times\int \,\mathrm{d}^3 \bm{k}_1 \,\mathrm{d}^3 \bm{k}_2\, 
   e^{-i \bmk_1 \cdot\left(\bm{\xi}_1 -\bm{\xi}_1'\right)-i \bmk_2 \cdot\left(\bm{\xi}_2 -\bm{\xi}_2^\prime\right)} \nonumber \\
   &&\times {\tau^{3\over2}\over (2\pi\sigma^2)^{9\over2}} \, \left({8\pi\sigma^2\over 2+\tau}\right)^{3\over2}
   \exp\left({2+\tau\over 8\sigma^2}t^2\right)\;
   \exp\left(-{1\over 16\sigma^2}\bigl(\alpha(\bm{\xi}_1+\bm{\xi}_1')^2+(\bm{\xi}_2+\bm{\xi}_2')^2
   \bigr)\right)\nonumber\\
   &&\times G_{3}(\bm p_1,\bm p_2,\bm p_3)|_{\bmP=\bmP_3} \ .
    \label{eq:main3}
\end{eqnarray}
Similar to Eq.~(\ref{eq:DeftandAlpha}), the following substitutions are applied 
\begin{equation} 
\label{eq:AlphaPrime}
{2+\tau\over 8\sigma^2}t^2-{\alpha\over 16\sigma^2}(\bm{\xi}_1+\bm{\xi}_1')^2 = -{\alpha'\over 16\sigma^2}(\bm{\xi}_1+\bm{\xi}_1')^2\ ,\qquad
\alpha'={\tau(2+\kappa)\over \kappa(2+\tau)}\ .
\end{equation}
This is useful to define a compact form of the expression, which allows for the evaluation of the probability of cluster formation as given below.
In the ase of the hypertriton, such probability reads
\begin{equation}
    \frac{\mathrm{d} N_{\hyp}}{\mathrm{d}^3 P} = {S_{3}\over (2\pi)^{12}} \int \mathrm{d}^3 \bm{k}_1 \,\mathrm{d}^3 \bm{k}_2\,
    G_{3}(\bm p_1,\bm p_2,\bm p_3)|_{\bmP=\bmP_3}
    \; {\cal P}(\bmk_1,\bmk_2,\sigma,\sigma')\ ,
\end{equation}
 where
\begin{eqnarray}
    {\cal P}(\bmk_1,\bmk_2,\sigma,\sigma')&=&
   {(\alpha')^{3\over2}\over 8(2\pi)^{3}\sigma^6} 
  \int \mathrm{d}^3 \bm{\xi}_1 \,\mathrm{d}^3 \bm{\xi}_2\int \mathrm{d}^3 \,\bm{\xi}_1^{\prime} \mathrm{d}^3 \bm{\xi}_2^{\prime}\; \varphi_{\hyp}^*(\bm{\xi}_1',\bm{\xi}_2')\varphi_{\hyp}(\bm{\xi}_1,\bm{\xi}_2) 
    \nonumber \\ 
    && \times 
   e^{-i \bmk_1 \cdot\left(\bm{\xi}_1 -\bm{\xi}_1'\right)-i \bmk_2 \cdot\left(\bm{\xi}_2 -\bm{\xi}_2^\prime\right)} \;  e^{-{1\over 16\sigma^2}\bigl(\alpha'(\bm{\xi}_1+\bm{\xi}_1')^2+(\bm{\xi}_2+\bm{\xi}_2')^2 \bigr)}\ .
    \label{eq:main4}
\end{eqnarray}
Eq.~\eqref{eq:main4} is finally used inside of ToMCCA. It is evaluated on a 4D $(|\textbf{k}_1|,|\textbf{k}_2|,\cos{\theta_{k_{12}}},\sigma)$ grid with a predetermined $\alpha'$. This evaluation is done once in advance, and the grid is saved in a 4D histogram.
As a test, ${\cal P}(\bm 0, \bm 0, \sigma\rightarrow 0,\sigma'=\sigma\sqrt{\tau})$ can be computed. In such a case, the Gaussians become delta functions, namely
\begin{equation}
 e^{-{1\over 16\sigma^2}\bigl(\alpha'(\bm{\xi}_1+\bm{\xi}_1')^2+(\bm{\xi}_2+\bm{\xi}_2')^2\bigr)} \rightarrow \left({16\pi \sigma^2\over \sqrt{\alpha'}}\right)^{3}
 \delta(\bm{\xi}_1+\bm{\xi}_1') \delta(\bm{\xi}_2+\bm{\xi}_2')\ ,  \qquad\sigma\rightarrow 0\ .
\end{equation}
Inserting this expression in Eq.~(\ref{eq:main4}), and applying the normalization condition of the wave function $\int d^3\xi_1 d^3\xi_2 |\varphi_{\lhThree}(\bm{\xi}_1,\bm{\xi}_2)|^2=1$, we
obtain
\begin{equation}
{\cal P}(\bm 0, \bm 0, 0,0)=64 \ .
\end{equation}
This is the expected value since the Wigner function $\mathcal{D}(0,0,0,0)=64$ by definition. This cross-check shows that the normalization as well as all prefactors are applied properly. 

The case of \heThree and \hThree can be recovered by setting $\kappa=1$ and $\tau=1$, from which follows that $\alpha'=1$ in Eq.~\eqref{eq:main4}. Details of the \heThree and \hThree wave functions can be found in Appendix~\ref{sec:hh}.
\subsection{Congleton approach for Hypertriton }
\label{sec:Congleton}
A simplified approach for calculating the $\lhThree$ yield has been developed employing the Congleton wave function~\cite{Congleton_1992}. This Congleton approach provides a relatively straightforward way to determine the $\lhThree$  wave function, assuming that the $\Lambda$ particle orbits an unperturbed deuteron within a $\Lambda$–d potential, $V_{\Lambda–\mathrm{d}}$. The potential is constructed from a separable N--$\Lambda$ interaction. This approach is justified since the $\Lambda$ separation energy of $B_\Lambda\approx130$ keV is only 6\% of the deuteron binding energy making it a true losely bound state. Furthermore, it has been shown~\cite{Bellini2021} that with a correct choice of parameters, the hypertriton $pn-\Lambda$ form factor calculated from pionless effective field theory ($\cancel{\pi}$EFT)~\cite{Hildenbrand2019Sep}.

The governing Hamiltonian for the system, including the center-of-mass motion, is given by
\begin{equation}
\begin{aligned}
H &= \frac{p^2}{2 m_\mu} + V_{\mathrm{np}} + \frac{q^2}{2 \mu} + V_{\Lambda-\mathrm{d}}\\
  &= H_{\text{d}}(\boldsymbol{p}) + V_{\Lambda}(\boldsymbol{q})\,.
\end{aligned}
\end{equation}
The Hamiltonian is split into two parts using a static approximation, in which the effective $\Lambda$--d system is considered. This way, the influence of the $\Lambda$ particle on the individual nucleons is neglected. Consequently, the $V_{\Lambda-\mathrm{d}}$ potential depends only on the $\Lambda$ momentum in the center of mass frame $\boldsymbol{q}$ (after accounting for spin dependence). Furthermore, the reduced mass of the $\Lambda$--d system is $\mu = 3.547 \,\mathrm{fm}^{-1}$ while the reduced mass of the two-nucleon system is $m_\mu = 2.379 \,\mathrm{fm}^{-1}$. The wave function can be expressed as the product of the spin function ($s =\frac{1}{2}$ for Hypertriton) and the nuclear wave function part
\begin{equation}
\begin{aligned}
\left\langle \boldsymbol{p}, \boldsymbol{q} \mid {}_{\Lambda}^3 H ; m_j \right\rangle = \psi_{\Lambda}(q) \sum_{(l, s) = \left(0, \frac{1}{2}\right), \left(2, \frac{3}{2}\right)} \psi_{\mathrm{d}}^{(l)}(p) 
\times \left[\mathcal{Y}_{l 0}^{\prime}(\hat{p}, \hat{q}) \otimes \chi_{1 \frac{1}{2}}^{\mathcal{S}}\right]_{\frac{1}{2} m_j} \frac{1}{\sqrt{2}} [\Lambda(\uparrow \downarrow - \downarrow \uparrow)]\,,
\end{aligned}
\end{equation}
where $\psi_{\Lambda}(q)$ and $\psi_{\mathrm{d}}^{(l)}(p)$ represent the $\Lambda$ wave function and the radial part of the deuteron wave function in the momentum space, respectively. The spin component describes the coupling between the spin-1 deuteron and the spin-$\frac{1}{2}$ $\Lambda$ particle, resulting in a total spin of either $\frac{1}{2}$ or $\frac{3}{2}$. The $s$-wave component of the deuteron corresponds to $S=\frac{1}{2}$, while the $d$-wave component corresponds to $S=\frac{3}{2}$. Apart from the spin component, the spatial wave function is a simple product of the deuteron and $\Lambda$ wave functions. Similar to the HH approach, the total wave function for the hypertriton $\Psi_{\hyp}\left(\bmx_1,\bmx_2,\bmx_3\right)$ can be factorized into a plane wave for the center-of-mass motion and an internal wave function that depends only on two relative coordinates, as shown in Eq.~\ref{eq:WavefunctionHyp}. The internal wave function is given as the product of $\Lambda$ and deuteron wave functions.
\begin{equation}
    \varphi_{{\lhThree}}(\bm{\xi}_1,\bm{\xi}_2) = \varphi_{\Lambda}(\bm{\xi}_1)\times\varphi_{\mathrm{d}}(\bm{\xi}_2)\,, 
\end{equation}
where, $\varphi_{\Lambda}(\bm{\xi}_1)$ and $\varphi_{\mathrm{d}}(\bm{\xi}_2)$ are the corresponding Fourier transform of the momentum space wave functions $\psi_{\Lambda}(q)$ and $\psi_{\mathrm{d}}^{(l)}(p)$ respectively. The final probability is given in terms of the Wigner functions for deuteron and for the $\Lambda$-d two-body system as
\begin{equation}
    \mathcal{P}( \bm k_{1}, \bm k_{2},\sigma)
    = \frac{\left(2 +\kappa^2\right)^{3/2}}{(2 +\kappa)^3}\ \frac{S_{\mathrm{\hyp}}}{(2\pi)^{3}\sigma^6}\,\,
   \int \mathrm{d}^3 \bm{r}_1 \int \mathrm{d}^3 \bm{r}_2\,\, \mathcal{D}_{\Lambda}(\bmk_{1},\bm{r}_{1})\mathcal{D}_d(\bm k_{2},\bm{r}_{2})\exp\left(-\frac{r_1^2(2 +\kappa)^2+4 r_2^2\left(2+\kappa^2\right)}{4 \sigma^2(2+\kappa)^2}\right).
\end{equation}
The Wigner densities of the $\Lambda$ particle denoted as $\mathcal{D}_{\Lambda}(\bmk_{1},\bm r_{1})$ is calculated in momentum space and the deuteron one, $\mathcal{D}_d(\bm k_{2},\bm r_{2})$, is computed in coordinate space. The computation of $\mathcal{D}_d(\bm k_{2},\bm r_{2})$ has already been provided in a previous work~\cite{Mahlein2023}, using the \AV wave function, while the derivation of $\mathcal{D}_{\Lambda}(\bmk_{1},\bm r_{1})$ is given in Appendix~\ref{App:Congleton}. The radial $\Lambda$ wave function in momentum space is then given by~\cite{Congleton_1992}
\begin{equation}
   \psi_{\Lambda}(q) = N(Q_\Lambda)\frac{\exp\left(-\left(\frac{q}{Q_\Lambda}\right)^2\right)}{q^2+\alpha_\Lambda^2},
    \label{eq:WFGaussNorm}
\end{equation}
where $N(Q_\Lambda)$ is a normalization constant
\begin{equation}
    N(Q_\Lambda) = \left\{\frac{\pi }{4 \alpha_\Lambda } \left[\left(\frac{4 \alpha_\Lambda ^2}{Q_\Lambda^2}+1\right) \text{Cerfe}\left(\frac{\sqrt{2} \alpha_\Lambda }{Q_\Lambda}\right)-\frac{2 \alpha_\Lambda  \left(\frac{2}{\pi }\right)^{1/2}}{Q_\Lambda}\right]\right\}^{-1/2}.
    \label{eq:Norm}
\end{equation}
The \texttt{Cerfe} function is defined as $\mathrm{Cerfe}(x)=\exp(x^2)(1-\mathrm{erf}(x))$. In the original publication~\cite{Congleton_1992} the parameter $\alpha_\Lambda=\sqrt{\mu B_\Lambda}=0.068$~fm$^{-1}$ was fixed to the binding energy and $Q_\Lambda=1.17$~fm$^{-1}$ was adjusted to the $\Lambda$ separation energy. In Ref.~\cite{Bellini2021}, the parameter $Q_\Lambda$ of the wave function was adjusted to match the form factor of Ref.~\cite{Hildenbrand2019Sep} obtained from $\cancel{\pi}$EFT. An excellent agreement is reached when $Q_\Lambda=2.5$~fm$^{-1}$ instead of $Q_\Lambda=1.17$~fm$^{-1}$. With these parameters $N(Q_\Lambda=2.5~\mathrm{fm}^{-1})^2=0.094369$~fm$^{-1}$ and $\int |\psi_{\Lambda}(q)|^2 \diff^3q =4\pi$. The final Wigner function $\mathcal{D}_{\Lambda}(\bm q,\bm r)$ is given as 
\begin{equation}
   \mathcal{D}_{\Lambda}(\bm q,\bm r) = N(Q_\Lambda)^2\int_0^{\infty} \diff k \frac{4 i \pi  \left(-1+e^{2 i k r}\right)}{q r \left(k^2+4 \left(\alpha_\Lambda ^2+q^2\right)\right)}\log \left(\frac{4\alpha_\Lambda ^2+(k-2 q)^2}{4 \alpha_\Lambda ^2+(k+2 q)^2}\right) \exp\left(-\frac{k^2+2 i k Q_\Lambda^2 r+4 q^2}{2 Q_\Lambda^2}\right),
    \label{eq:WigLambda}
\end{equation}
and $\mathcal{D}_{\Lambda}(0,0)/4\pi = 8.0$ as expected given the normalization of the wave function. 
\section{ToMCCA upgrade for $A=3$ coalescence}
\label{sec:ToMCCAExtension}
For the extension of the coalescence predictions, the ToMCCA model~\cite{Mahlein2024} had to be upgraded to incorporate the production of $A=3$ nuclei. The first step is to include an additional particle loop. For \heThree this loop adds another proton, looping over $N_\mathrm{p}-1$ nucleons. Due to the isospin symmetry assumed in ToMCCA, the case of \hThree is also covered since the first and third particles can be assumed to be neutrons. The case of \lhThree is slightly more complicated. Here, the third loop includes a $\Lambda$, which no longer abides by the isospin symmetry. The measured $\Lambda$ spectra from the ALICE collaboration~\cite{LambdaProtonALICE} are parameterized similarly to the protons and included. The spectra are characterized using a Levy-Tsallis function
\begin{equation}
    \frac{\diff^2N}{\diff y\diff \pt}=\frac{\pt\frac{\diff N}{\diff y}(n-1)(n-2)}{nT(nT+m_\Lambda(n-2))}\left(1+\frac{\sqrt{m_\Lambda^2+\pt^2}-m_\Lambda}{nT}\right)^{-n}\ ,
\end{equation}
where $m_\Lambda=1.115\mathrm{~GeV}/c^2$ is used and the parameters are described as a function of the mid-rapidity multiplicity \dNetaOFive\footnote{For sake of readability \dNetaOFive$\rightarrow N_\mathrm{ch}$} as follows
\begin{align}
    \frac{\diff N}{\diff y}&=\frac{N_\mathrm{ch}^2}{0.7597+N_\mathrm{ch}^2}0.03646N_\mathrm{ch}\\
    T&=0.08306N_\mathrm{ch}^{0.5094}+0.04290\\
    n&=0.3050N_\mathrm{ch}+6.331,
\end{align}
and the resulting $\Lambda$-spectra from ToMCCA are compared to the measurement in Fig.~\ref{fig:LambdaSpectra}.

The second major change is the used phase space of nucleons. Previously, ToMCCA provided an uncorrelated phase space, including only very basic correlations between particles, which were mainly driven by angular correlations. However, fully-fledged Monte Carlo generators include more detailed correlations, especially between the spatial position and momentum of each particle. In order to incorporate these into ToMCCA, the EPOS~3\footnote{The EPOS~3.117 event generator was used with the special setting to force the impact parameter $b=0$.} event generator was utilized, and the two-particle phase space was extracted. This means that for each nucleon-nucleon pair in the EPOS simulation, the relative momentum $k^*$, distance, and \mmt were extracted and stored. Further, each event was classified by the mid-rapidity multiplicity \dNetaOFive. This provides a 4-dimensional distribution encoding the entire two-particle correlations provided by EPOS. In ToMCCA, each pair draws its source size from this distribution once $k^*$, $N_\mathrm{ch}$ and \mmt are determined. The phase space is assumed to be equal for nucleon-nucleon and nucleon-hyperon pairs. This leaves three different source sizes for the nucleon triplet. These three sizes are averaged according to 
\begin{equation}
    R_\mathrm{av}=\sqrt{\frac{8(R_1^2+R_2^2+R_3^2)(R_2^2R_3^2+R_1^2(R_2^2+R_3^2))^2}{3R_1R_2R_3\sqrt{R_1^{-2}+R_2^{-2}+R_3^{-2}}((R_1^2+R_2^2)(R_1^2+R_2^2+4R_3^2))^{5/2}}},
\end{equation}
which has been obtained by comparing the mean of the Gaussian source distribution of three equal-sized sources to one of three independent sources. This change of the phase space also requires a refitting of the source size\footnote{The compatibility of this new parameterization with ALICE $\sqrt{s}=5$ TeV pp data is shown in Fig.~\ref{fig:DSpectra}.}. The source size scaling with \mmt has been previously parameterized as a power law
\begin{eqnarray}
\label{eq:SourceParameterization}
    \sigma(\mmt)=B(\mmt)^{-C}.
\end{eqnarray}
The parameters $B$ and $C$ also depend on $N_\mathrm{ch}$ as

\begin{align}
    B(N_\mathrm{ch})&=\frac{0.7359N_\mathrm{ch}^{1/3}}{1+e^{-0.1808(N_\mathrm{ch}-12.19)}}-\frac{1.945}{1+e^{-0.1808(N_\mathrm{ch}-12.19)}}+1.127,\\
    C(N_\mathrm{ch})&=-\frac{0.2218}{0.3168+e^{-0.1834N_\mathrm{ch}-5.571}}+0.9757.
\end{align}
Similar to the previous parameterization~\cite{Mahlein2024}, these source radii are not to be understood as a genuine prediction but an effective parameterization. 
Lastly, the angular correlations need to be extended to the three-particle case. The base assumption is that two-particle correlations are dominant over genuine three-particle correlations. Then, the relative angular distribution can be constructed from two-particle correlations. The first particle is assigned an absolute azimuthal value from a flat distribution. Then, from the two-particle correlations used in the $A=2$ case, the second particle is assigned an angular distribution
\begin{equation}
    \mathcal{P}(\varphi_2)=1\times(N\sin(a\Delta\varphi_{12}-b)+c),
\end{equation}
where $\Delta\varphi_{12}=\varphi_1-\varphi_2$ and $a=1,b=\pi/2$. The parameters $c$ and $N$ depend on $N_\mathrm{ch}$ like
\begin{align}
c=&0.8095+0.1285\cdot N_\mathrm{ch}^{0.1114}    \\
N=&0.4061\cdot N_\mathrm{ch}^{-0.3725}.
\end{align}
These parameters were obtained from fitting the $\Delta\varphi$ correlation functions as a function of multiplicity obtained by the ALICE collaboration~\cite{ALICEDeltaPhi}. The third particle undergoes the same procedure, but this time depending on two relative angles $\Delta\varphi_{13}$ and $\Delta\varphi_{23}$, giving
\begin{equation}
    \mathcal{P}(\varphi_3)=1\times(N\sin(a\Delta\varphi_{13}-b)+c)\times(N\sin(a\Delta\varphi_{23}-b)+c).
\end{equation}

\section{Comparison to experimental data}
\subsection{\heThree and \hThree production}
\begin{figure}[!hbt]
    \centering
    \subfigure{\centering \includegraphics[width=0.49\linewidth]{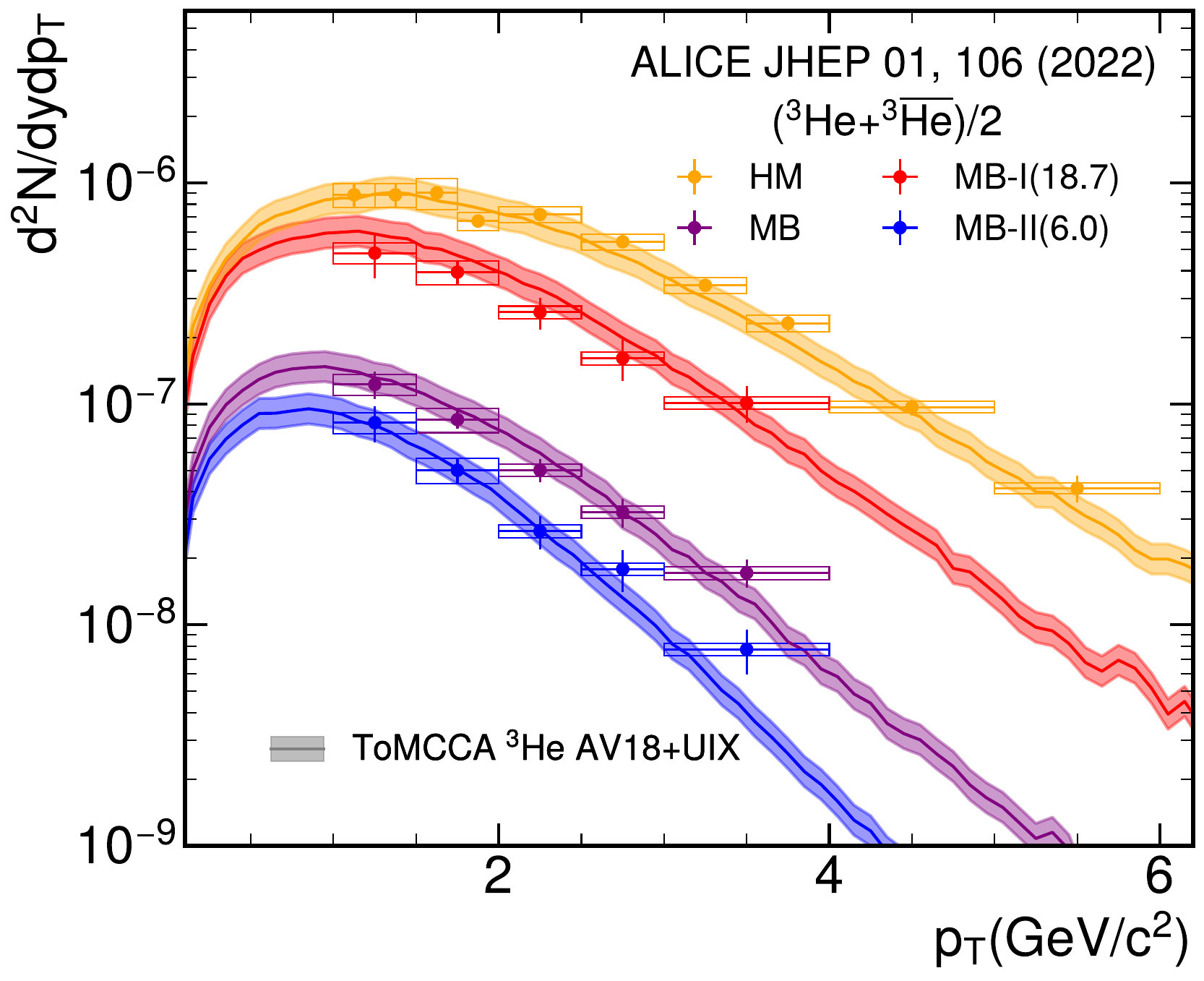}}
    \subfigure{\centering \includegraphics[width=0.49\linewidth]{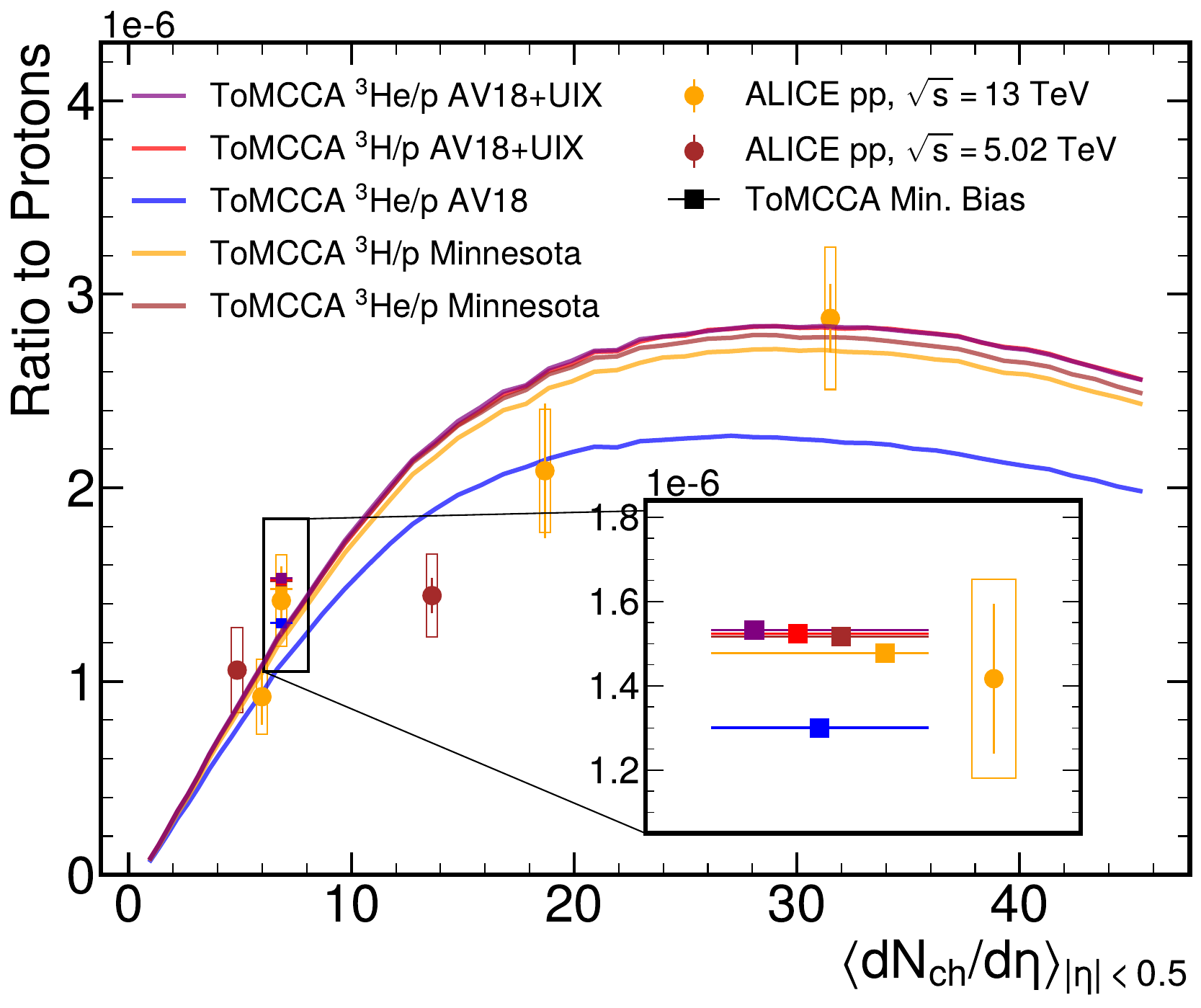}}  
    \caption{(left) The \heThree \pt spectra measured by ALICE~\cite{ALICENuclei13TeV} in pp collisions at $\sqrt{s}=13$~TeV in four intervals of multiplicity. The corresponding predictions for ToMCCA are shown as the colored bands. The width of the central line indicates the statistical uncertainty, while the shaded band shows the systematic uncertainties of $\pm17\%$ (see Sec.~\ref{sec:Uncertainties}). (right) The \heThree/p (\hThree/p) ratio as a function of \dNetaOFive and comparison to the ALICE measurements~\cite{ALICENuclei13TeV,ALICE:2021ovi}. Different assumptions for the wave function are tested, only 2-body forces based on the \AV potential, 2-body+3-body \AV+UIX wave function for \heThree and \hThree, as well as wave function based on the 2-body Minnesota potential. The predictions and the ALICE measurement for 13 TeV minimum bias collisions are shown as the squares and circles, magnified in the inlet. For the sake of visibility the points are shifted on the x-axis.}
    \label{fig:He3Spectra}
\end{figure}

In this section the output of ToMCCA is compared to experimental data. The AV18+UIX wave function is considered as a reference, but also other wave functions are tested and differences are discussed. These wave functions are calculated using the PHH method~\cite{Kievsky:2008es}. Figure~\ref{fig:He3Spectra} (left) shows the \heThree \pt spectra measured by ALICE~\cite{ALICENuclei13TeV} in four different \dNetaOFive intervals: i) high multiplicity (HM, blue) with \dNetaOFive$=31.5\pm0.3$, ii) minimum bias\footnote{INEL>0 by ALICE definitions.} (MB, orange) with \dNetaOFive$=6.9\pm0.1$, and a subdivision of minimum bias into iii) MB-I (red) and iv) MB-II (purple), with \dNetaOFive$=18.7\pm0.3$ and $6.0\pm0.2$ respectively. The shown experimental uncertainties are statistical (errorbars) and systematic (boxes). For the minimum bias predictions, the published multiplicity distributions~\cite{ALICEMultiplicity} were used (See Sec.~\ref{sec:MinBias}) instead of the Erlang functions used for the multiplicity-dependent studies \cite{Mahlein2024}. The predictions from ToMCCA using the AV18+UIX wave function are shown as colored lines. Figure~\ref{fig:He3Spectra} (right) shows the integrated yield ratios \heThree/p and \hThree/p as a function of \dNetaOFive. Shown are the measurements by the ALICE collaboration~\cite{ALICENuclei13TeV,ALICE:2021ovi} in pp collisions at $\sqrt{s}=5$~TeV and $\sqrt{s}=13$~TeV. Shown are the systematic uncertainties as boxes and statistical ones as errorbars. The colored lines indicate the predictions from ToMCCA using different assumptions of the wave function. The width of the line indicates the statistical uncertainty of the model. In the inlet the minimum bias (INEL>0) measurement is compared to the model predictions. Importantly, these are not to be compared to the multiplicity continuum lines, since minimum bias collisions have a very wide multiplicity distribution.
While all predictions with the correct binding energy (AV18+UIX or Minnesota) reproduce the data at 13~TeV well, there seems to be a tension in the higher-multiplicity measurement at $\sqrt{s}=5$~TeV. Overall, the model reproduces all measurements within 2 standard deviations, when considering the model uncertainties (see Sec.~\ref{sec:Uncertainties}). Measurements in intermediate mass collision systems, such as O--O and Ne--Ne, could help rule out the two-body only hypothesis for the wave function.

\begin{figure}[!hbt]
    \centering
    \subfigure{\centering \includegraphics[width=0.49\linewidth]{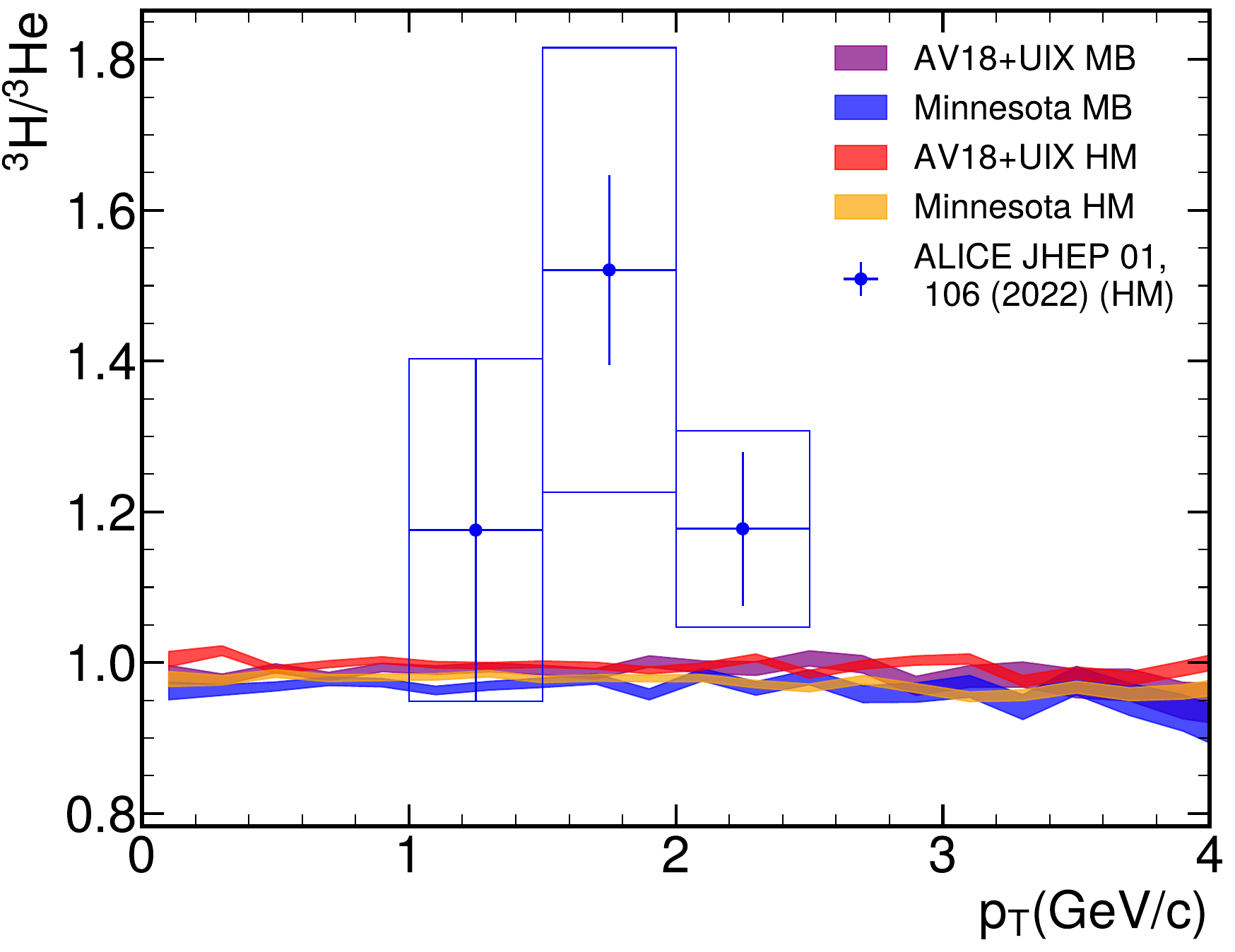}}
    \subfigure{\centering \includegraphics[width=0.49\linewidth]{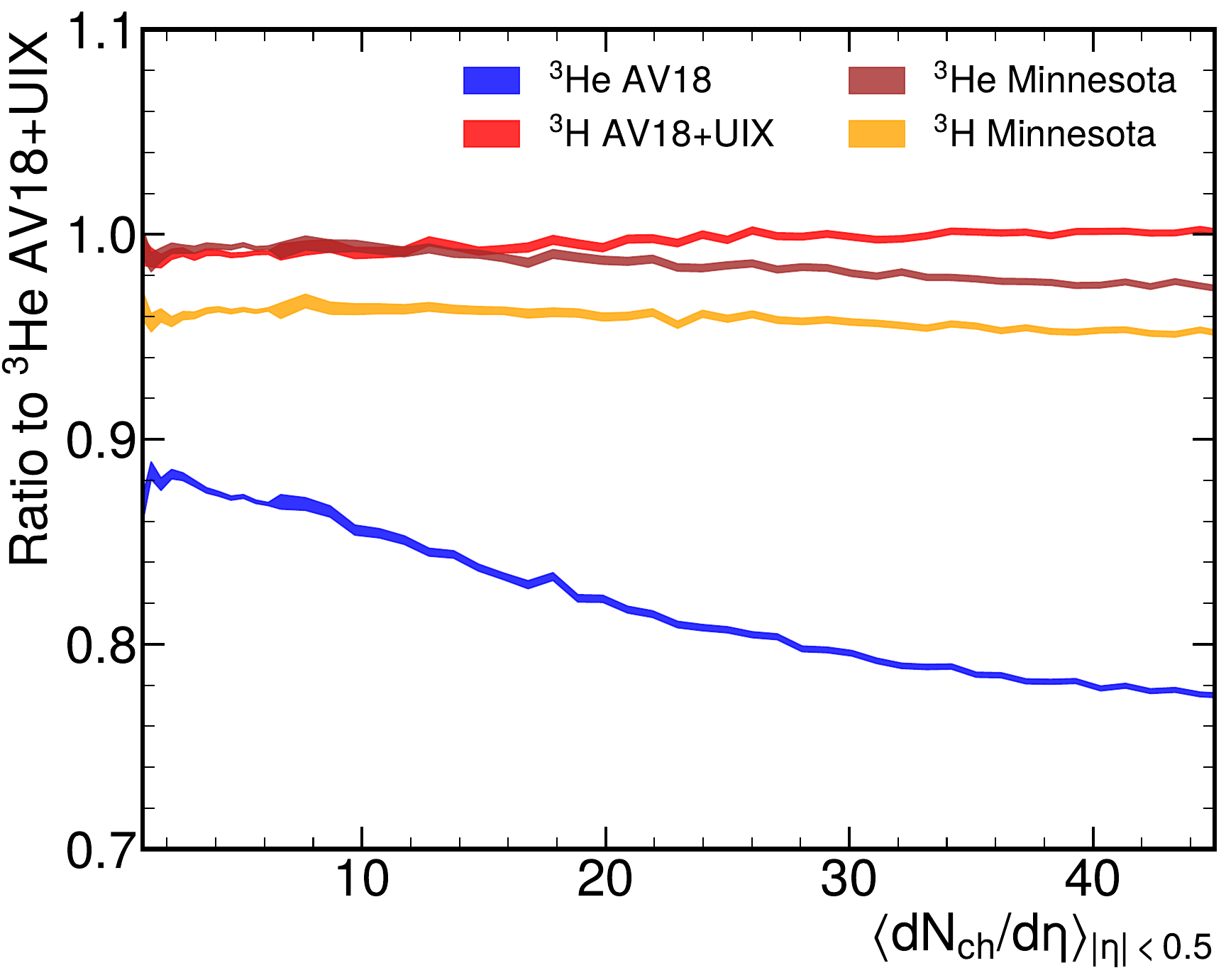}}  
    \caption{(left) \hThree/\heThree ratio as a function of \pt predicted by ToMCCA for high multiplicity (HM, \dNetaOFive=31.5) and minimum bias (MB, \dNetaOFive=6.9) collisions, using two different assumptions on the wave function (AV18+UIX and Minnesota). They are compared to the measurement by ALICE in HM pp collisions at $\sqrt{s}=13$~TeV. All predictions by ToMCCA are compatible with unity within 5\% and with the measurement within two standard deviations. (right) Ratio of the ToMCCA predictions using several wave function assumptions to the $\heThree$ one using AV18+UIX.}
    \label{fig:HeTRatio}
\end{figure}

Reference~\cite{Sun2019} predicted that there should be a difference in production yield between \heThree and \hThree due to their different size ($r_\mathrm{^3He}/r_\mathrm{^3H}\approx1.11$), which gets more pronounced towards lower \dNetaOFive corresponding to smaller source radii. The interplay between the size of the nucleus ($r_{\heThree}=1.77$ fm, $r_{\hThree}=1.6$fm) and the source size ($\sigma\approx1-1.5$ fm) causes a suppression of larger nuclei for smaller source sizes. The ToMCCA prediction of the yield ratio between \hThree and \heThree as a function of \dNetaOFive (right) and as a function of \pt (left) is shown in Fig.~\ref{fig:HeTRatio}. Interestingly, ToMCCA predicts no dependence of this ratio on the source size, showing instead a constant ratio of $0.996\pm0.004$ over the whole multiplicity range, with a small deviation from unity and a ratio of $0.991\pm0.002$ from \dNetaOFive$\in[0,10]$. The main difference between the two predictions are the nuclear wave functions. While Ref.~\cite{Sun2019} uses Gaussian wave functions, ToMCCA uses more realistic assumptions on the structure of the nuclei. Furthermore, previous work has focused on reproducing the charge (or proton) radii of nuclei rather than the mass radii. Since \heThree is doubly charged, its charge radius naturally increases. From AV18+UIX calculations, one can obtain the mass radii for both nuclei, and they are equal within 2\% ($r_{M,^3\mathrm{H}}=1.683$ fm, $r_{M,^3\mathrm{He}}=1.715$ fm, see Tab.~\ref{tab:b3} for details.)
Figure~\ref{fig:HeTRatio} (right) also shows the effect of the 3-body force on the \heThree yield. The ratio of the \heThree yields between AV18, which considers only the 2-body forces, and AV18+UIX, which considers 2 and 3-body forces, is close to 0.9 for very small \dNetaOFive, with an increasing deviation towards higher multiplicities, up to 22\% at $\dNetaOFive=45$. This indicates an enhancement of the yield when attractive 3-body forces are included, reducing the size of the nucleus. The effect even gets more pronounced at larger multiplicities, corresponding to larger source sizes. This goes against the findings of Ref.~\cite{Sun2019}, where the yield of the smaller nucleus (e.g. \hThree) compared to the larger nucleus (e.g. \heThree) is enhanced at low multiplicities and smaller source sizes. 
Contrary to common wisdom, the size ultimately seems not to be the driving factor in nuclear production but the underlying interactions between the nucleons. Indeed, the UIX potential is mainly driven by long-range three-nucleon two-pion exchange. At larger sources, the long-range effects of the interaction become more important, and the attractive nature of the 3-body force increases the yields. Studies with other 3-body forces with varying ranges could be performed to quantify the effect of the interplay between source size and 3-body interaction range. This is relegated to future studies.
Another realistic assumption for the nuclear wave function is given by the Minnesota potential~\cite{MinnesotaPotential}, a simplified version of the nucleon-nucleon interaction (it does not include any tensor component). Despite its simplicity and the absence of a 3-body component, it is able to reproduce the binding energy of \heThree and \hThree within 2\%. The ratio to the AV18+UIX wave function predictions shows that this potential is also equally able to reproduce the yields, with a ratio of $0.984\pm0.016$ in the case of \heThree and $0.960\pm0.027$ in the case of \hThree. Interestingly, the \hThree yield is systematically lower by $\sim(2.7\pm0.3)\%$. This might be explained by the worse description of the binding energy for \hThree ($\approx1.2\%$ lower) compared to \heThree ($\approx0.1\%$ lower), as shown in Tab.~\ref{tab:b3}. 

\subsection{\lhThree production}
\begin{figure}[!hbt]
    \centering
    \subfigure{\centering \includegraphics[width=0.49\linewidth]{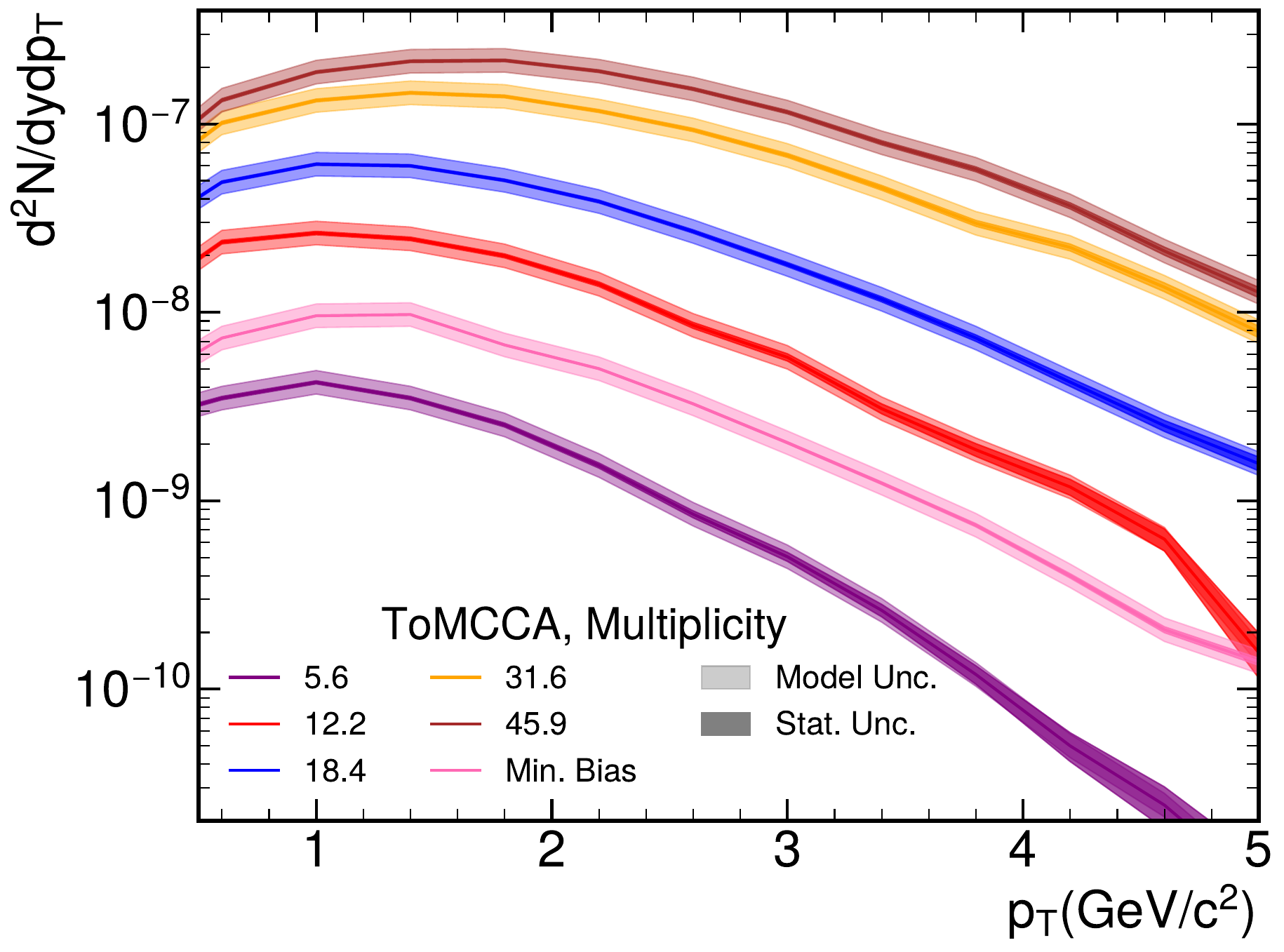}}
    \subfigure{\centering \includegraphics[width=0.475\linewidth]{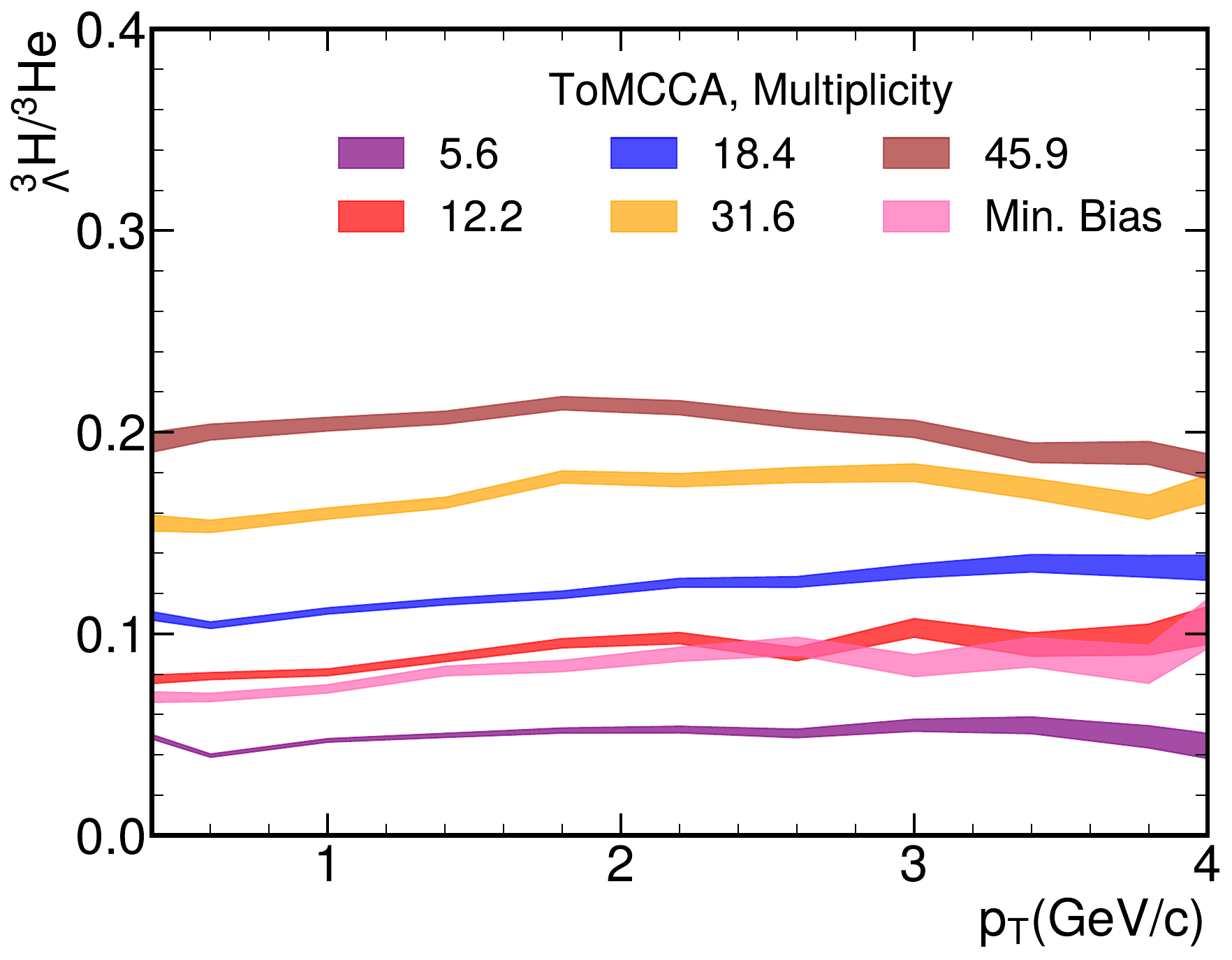}}  
    \caption{(left) The \lhThree \pt spectra using the Congleton wave function predicted by ToMCCA for various multiplicities from \dNetaOFive=5.6 to \dNetaOFive=45.9 as well as minimum bias collisions. No experimental data is available for comparison at this time. The light shaded bands indicate the model uncertainties (see Sec.~\ref{sec:Uncertainties}), the dark shaded bands represent statistical uncertainties. (right) Ratio of \lhThree/\heThree as a function of \pt for various multiplicity classes as well as minimum bias collisions. as predicted by ToMCCA using the Congleton wave function for \lhThree and \AV+UIX for \heThree. The shaded bands indicate the combined statistical uncertainties. No model uncertainties are estimated for these ratios.}
    \label{fig:LH3Spectra}
\end{figure}

\begin{figure}[!hbt]
    \centering
    \subfigure{\centering \includegraphics[width=0.49\linewidth]{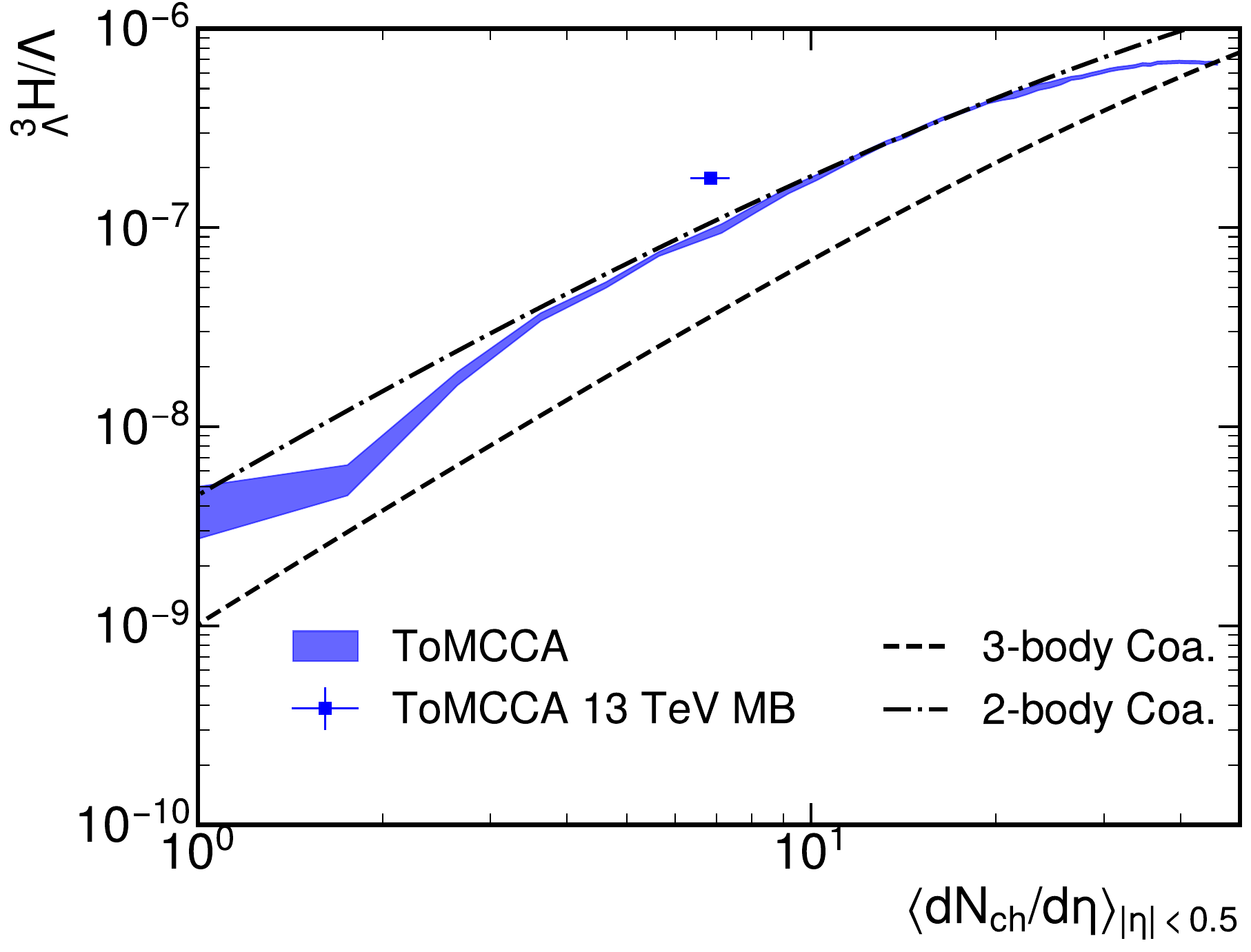}}
    \subfigure{\centering \includegraphics[width=0.49\linewidth]{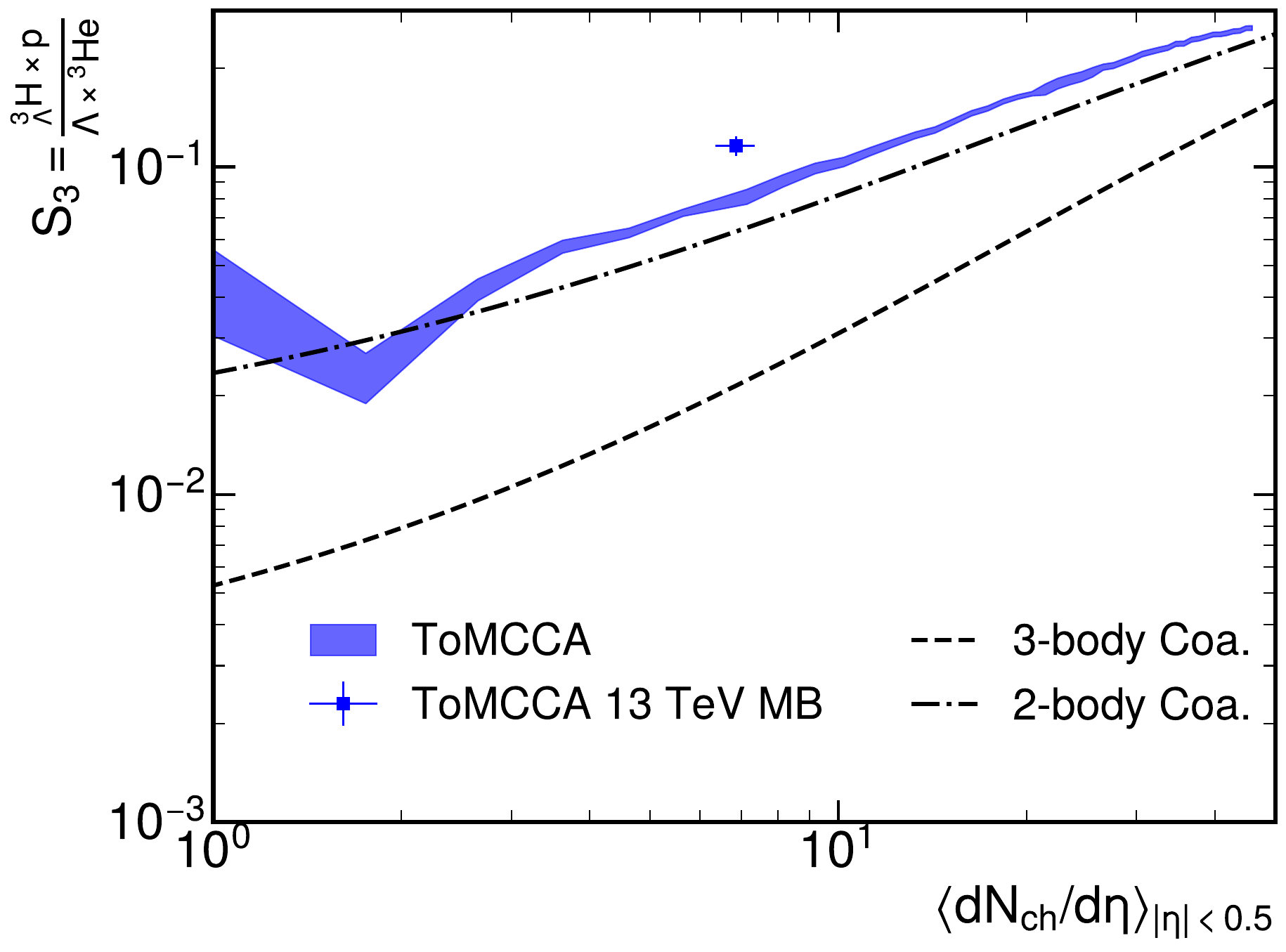}}  
    \caption{(left) \lhThree/$\Lambda$ ratio predicted by ToMCCA using the Congleton wave function are shown as the purple band. The dashed and dot-dashed lines are predictions from~\cite{Sun2019} using different assumptions for a Gaussian wave function. (right) The purple band shows the predictions of the $S_3$ parameter by ToMCCA using the Congleton wave function. The dashed and dot-dashed lines are the same model as on the left~\cite{Sun2019}.}
    \label{fig:LH3Ratios}
\end{figure}

ToMCCA is extended to include hyperon production as described in Sec.~\ref{sec:ToMCCAExtension}. For the \lhThree,  the Congleton wave function in combination with the \AV wave function is used (see Sec.~\ref{sec:Congleton}). Figure~\ref{fig:LH3Spectra} (left) shows the \lhThree \pt spectra for pp collisions for various multiplicities between \dNetaOFive=5.6 and \dNetaOFive=45.9 as well as minimum bias collisions. No published spectra are available for comparisons at this time for pp collisions. Figure~\ref{fig:LH3Spectra} (right) shows the \lhThree/\heThree ratio as a function of \pt. This observable is sensitive to differences between thermal production models and coalescence models. Thermal models usually determine the \pt spectra using blast-wave parameterizations, mimicking radial flow effects. Radial flow causes heavier particles to have harder spectra, which causes this ratio to rise with \pt. On the other hand, in coalescence models, the driving factor is the source size. At larger \pt a smaller source size is observed. This is supposed to cause suppression of larger nuclei such as \lhThree, and hence the ratio is suppressed. It is unclear how this observation stands compared to the \hThree/\heThree ratio, where the size of the nuclei did not influence the result. Indeed, in ToMCCA, these ratios are almost independent of \pt.
Figure~\ref{fig:LH3Ratios} (left) depicts the ratio of \lhThree/$\Lambda$ as a function of \dNetaOFive. Similarly to the \pt spectra, no integrated yields have been published for pp collisions yet. Only one measurement from p-Pb collisions has been published, however, it is unclear whether the transition from pp to p--Pb should be a smooth one. The ToMCCA prediction is shown as the blue band. The blue square marker shows the result of using the minimum bias multiplicity distribution, as used for \heThree. The yield is $\approx2$ times larger than using the Erlang distribution for \dNetaOFive$=6.9$, due to the much larger width of the distribution. Since coalescence is a non-linear process, i.e., it does not depend on the number of nucleons but the number of nucleon pairs, it is sensitive to higher moments of the multiplicity distribution. The right side of Fig.~\ref{fig:LH3Ratios} shows the strangeness population factor $S_3=\frac{^3_\Lambda\mathrm{H}\times\mathrm{ p}}{^3\mathrm{He}\times\Lambda}$. For the \heThree predictions, the AV18+UIX was used. The predictions from Ref.~\cite{Sun2019} are shown as the dashed and dot-dashed lines. These predictions use Gaussian wave functions for both \lhThree and \heThree, but \lhThree is either described by a homogenous wave function (\textit{3-body coalescence}) or a d-$\Lambda$ system (\textit{2-body coalescence}). The \heThree is described by \textit{3-body coalescence} in both cases.
\subsection{Model uncertainties}
\label{sec:Uncertainties}
In this section, the global yield uncertainties will be discussed. From the $A=2$ case~\cite{Mahlein2024}, the uncertainty on the source size, caused by the measured deuteron and proton spectra uncertainty, was 5.7\%. It was obtained by varying the $B$ parameter in Eq.~\eqref{eq:SourceParameterization} within the uncertainties obtained from fitting the deuteron spectra and is thus independent of \mmt. When varying the source size by this amount, the $^3$He yield changed by $\pm13\%$ for \dNetaOFive$=2$ and by $\pm18\%$ for \dNetaOFive$=45$, giving on average a variation of $\pm15.5\%$. In order to ensure numeric stability, the relative momenta of the considered nucleons cannot exceed $k_c=1$ GeV/$c$. In order to gauge the effect of this cutoff, its value for the \heThree and \hThree AV18+UIX wave functions was varied by $\pm50$ MeV, which resulted in a global variation of $\pm7\%$. Adding these two, a global uncertainty of $\pm17\%$ is reached. In the case of \lhThree, the cutoff variation can be omitted since the solution is numerically much more stable, and a global uncertainty of $\pm15.5\%$ is taken.

\section{Summary and Outlook}
In this paper, we have presented an upgrade of the ToMCCA model to incorporate coalescence for $A=3$ nuclei, including \heThree, \hThree, and \lhThree. The results for \heThree and \hThree are obtained for various different wave functions based on \AV for the 2-body potential and Urbana IX for the genuine 3-body potential, as well as the Minnesota potential, which is a 2-body potential capable of reproducing the binding energies without additional 3-body potentials. These wave functions are obtained by using the pair-correlated hyperspherical harmonics (PHH) approach. The results for all wave functions that reproduce the binding energies are compatible with each other, and only the \AV without UIX wave function shows a significant deviation. Interestingly, the \hThree/\heThree ratio shows no dependence on the charged-particle multiplicity, which is in contrast to previous studies using Gaussian wave functions~\cite{Sun2019}. Overall, the description of \heThree yields and their \pt spectra is in agreement with the measurements by ALICE in pp collisions at $\sqrt{s}=13$~TeV. In the case of \lhThree the Congleton approach~\cite{Congleton_1992} for the wave function was tested. It includes an undisturbed deuteron, described using the \AV wave function, and a quasi-free $\Lambda$ around it. With this wave function, predictions for the yields and \pt spectra in pp collisions are presented. In a future study, the HH description of \lhThree can be tested, which can also be extended to the elusive $\Lambda$nn state. As of now, no data on \lhThree production in pp collisions has been published. The \lhThree/\heThree ratio was previously predicted to drop off with increasing \pt, which is used a common differentiator between thermal production and coalescence models. The ToMCCA predictions show no such behaviour within the uncertainties. The minimum bias predictions of the \lhThree yield are about 2 times higher than the yield at the same mean multiplicity, but with a narrow multiplicity interval using Erlang functions.

With this successful upgrade of the ToMCCA model to $A=3$ coalescence, it can be used for predictions of nuclei yields at different collision energies, specifically the ones interesting for indirect dark matter searches. Furthermore, the predictions could help in determining the production mechanism of light nuclei once high-precision data from the LHCs Run-3 and Run-4 campaigns become available.

\vspace{1cm}
\noindent
\textbf{Acknowledgments}
We gratefully acknowledge the support of the INFN-Pisa computing center.
\section*{Declarations}
This work is supported by the European Research Council (ERC) under the European Union’s Horizon 2020 research and innovation programme (Grant Agreement No 950692) and the BMBF 05P24W04 ALICE. The work of Bhawani Singh was partially supported by the U.S. Department of Energy, Office of Science, and Office of Nuclear Physics under contract DE-AC05-06OR23177.

\appendix

\section{Deuteron Spectra with new source parameterization}
\begin{figure}[!hbt]
    \centering
    \includegraphics[width=0.99\linewidth]{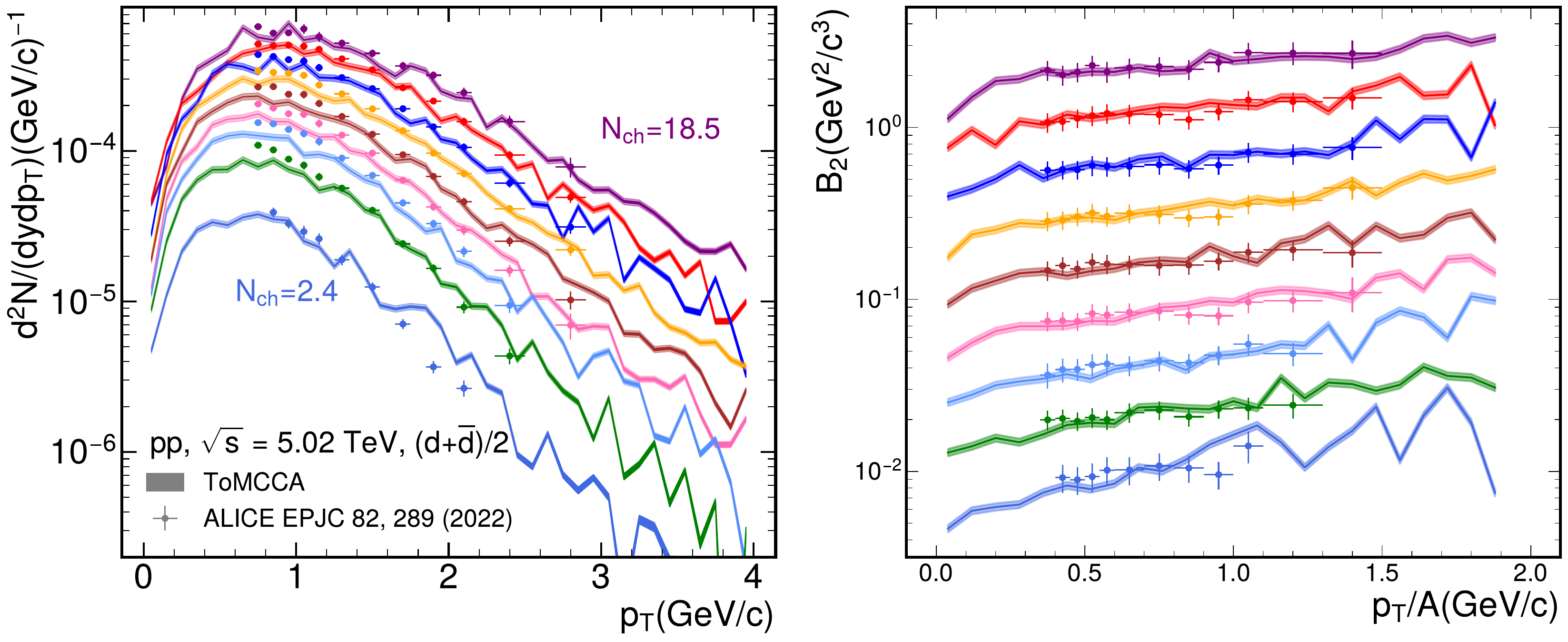}
    \caption{(left) Comparison of ToMCCA with ALICE deuteron ($(d+\overline{d})/2$)-spectra using the new parameterization of the source and the EPOS phase space. (right) Comparison of $B_2$ predictions by ToMCCA with the measurements of ALICE for pp collisions at $\sqrt{s}=5$~TeV. The colored bands indicate the previously determined 4.6\% uncertainty on the predicted deuteron spectra, stemming from the experimental uncertainty of the proton spectra.}
    \label{fig:DSpectra}
\end{figure}
Since the source is retuned using the EPOS $q-r$-phase space as input, its performance with deuterons needs to be reassessed. The main change comes from the fact that particles with small relative momenta also have relatively smaller distances, thus boosting coalescence probability. Overall, the deuteron spectra are well described by the newly tuned model within the experimental and model uncertainties, as shown in Fig.~\ref{fig:DSpectra}.

\section{$\Lambda$ spectra and parameterization}
\begin{figure}[!hbt]
    \centering
    \subfigure{\centering \includegraphics[width=0.49\linewidth]{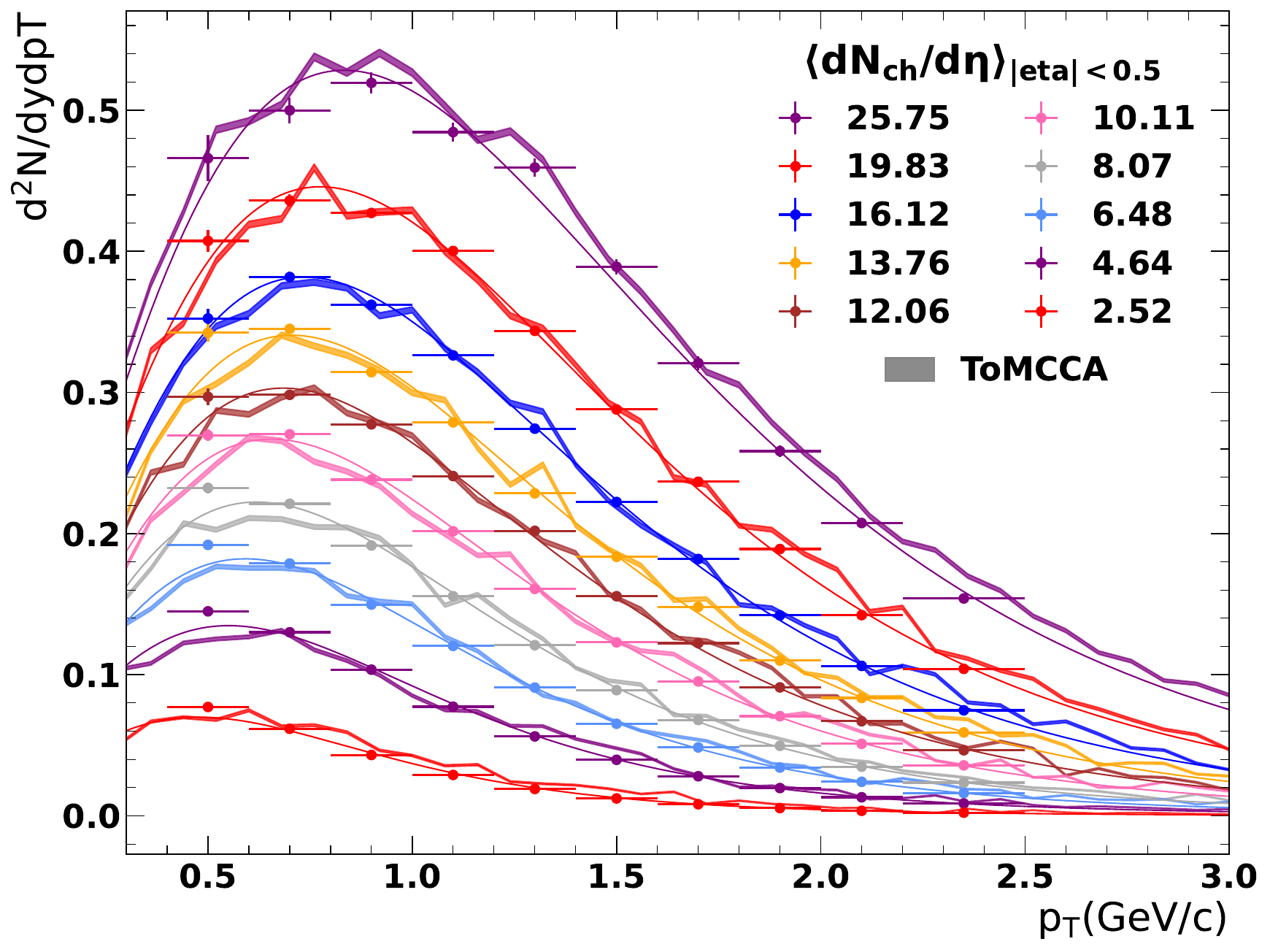}}
    \subfigure{\centering \includegraphics[width=0.49\linewidth]{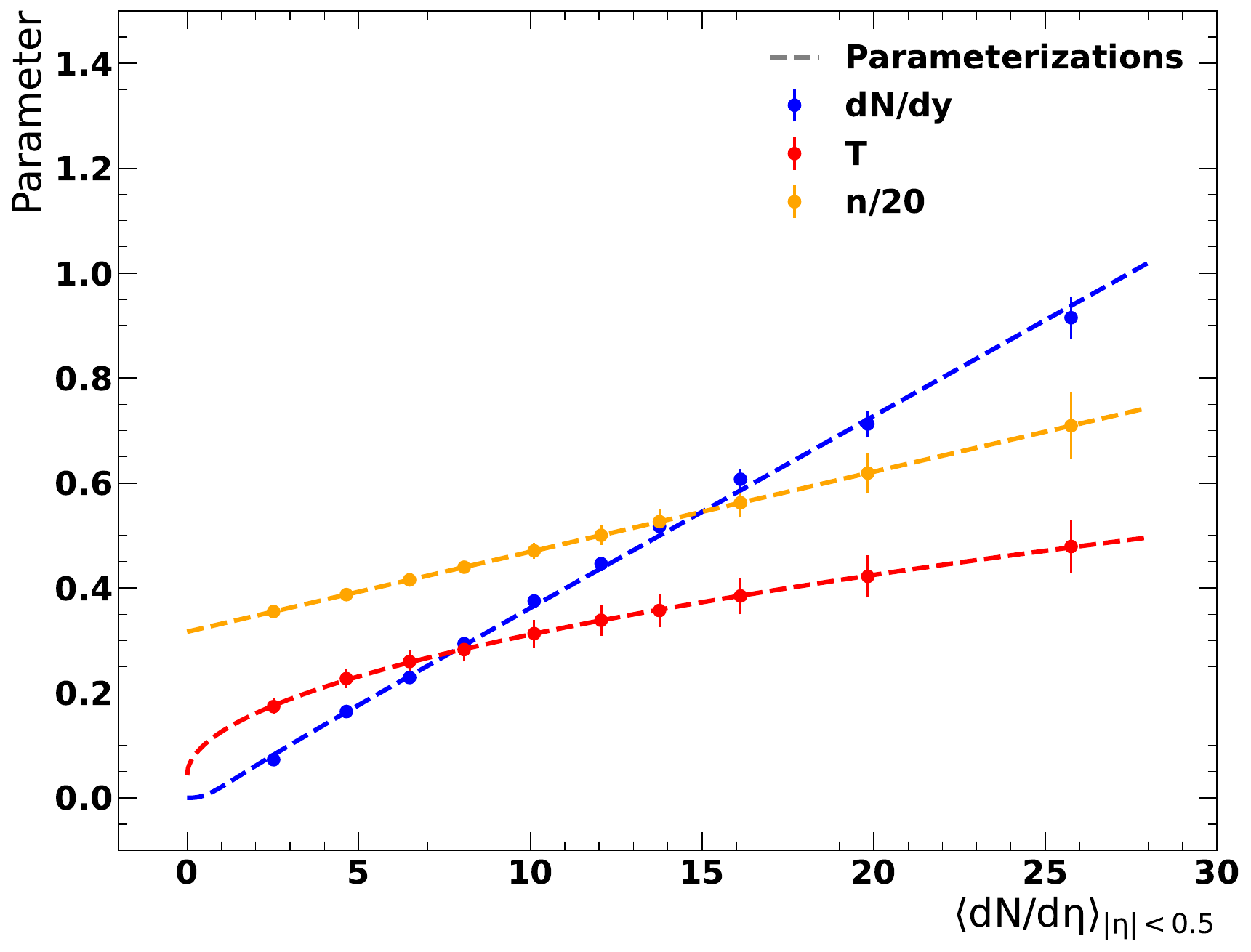}}  
    \caption{(left)$\Lambda$ \pt spectra measured by ALICE in pp collisions at $\sqrt{s}=13$~TeV~\cite{LambdaProtonALICE} compared to the ToMCCA predictions with above parameterization. (right) Extracted fit parameters as a function of \dNetaOFive and the corresponding fit functions.}
    \label{fig:LambdaSpectra}
\end{figure}

Fig.~\ref{fig:LambdaSpectra} (left) depicts the $\Lambda$ spectra measured by ALICE~\cite{LambdaProtonALICE} in pp collisions at $\sqrt{s}=13$~TeV in 10 multiplicity classes between \dNetaOFive=2.5 and \dNetaOFive=25.75. The ToMCCA predictions are shown alongside the Levy-Tsallis fit to the experimental data. The parameters of this Levy-Tsallis fit are shown in the right panel for the various multiplicity classes. Overall a good agreement with the measurements is achieved.
\section{$B_3$ parameter}
\begin{figure}[!hbt]
    \centering
    \subfigure{\centering \includegraphics[width=0.49\linewidth]{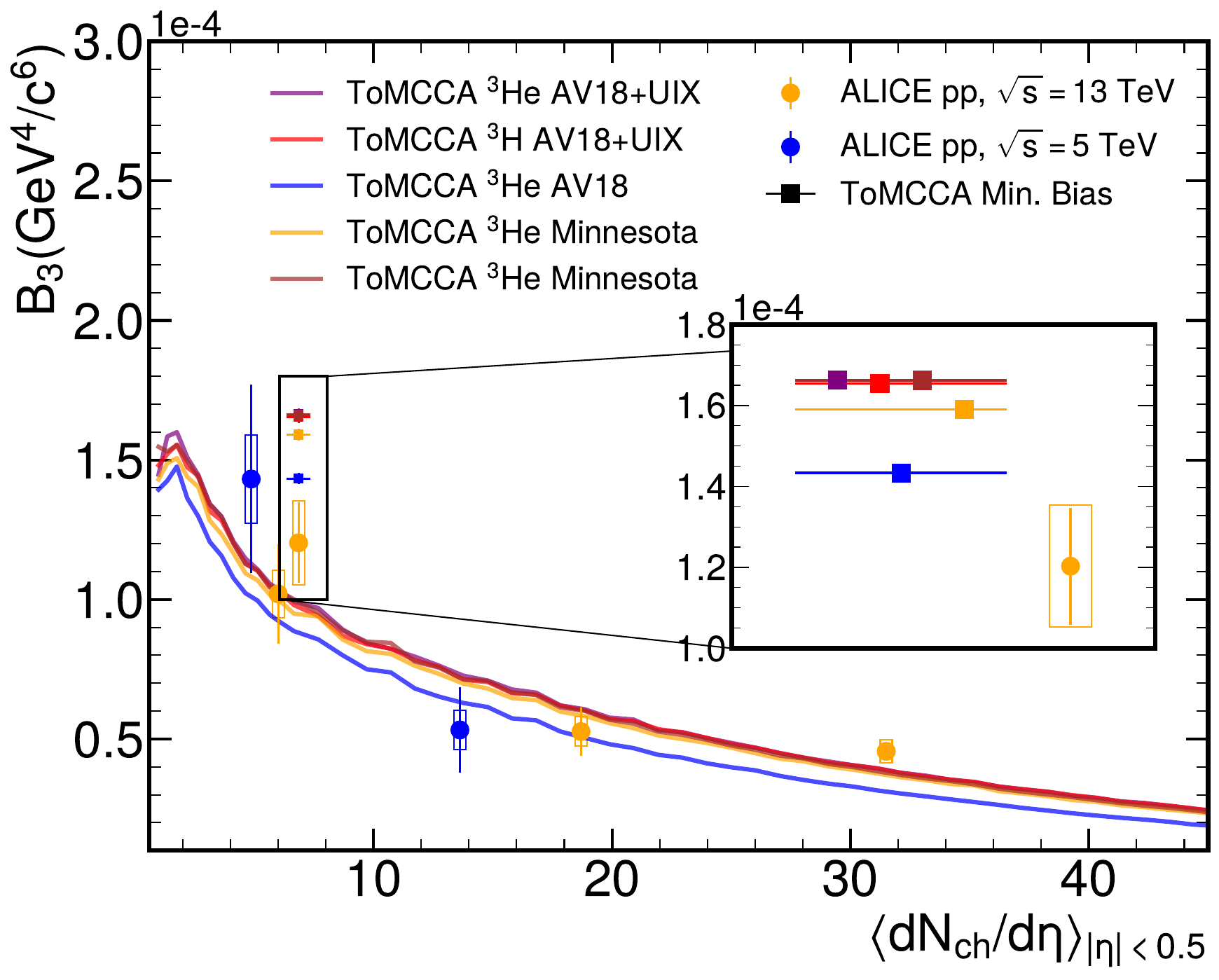}}
    \subfigure{\centering \includegraphics[width=0.49\linewidth]{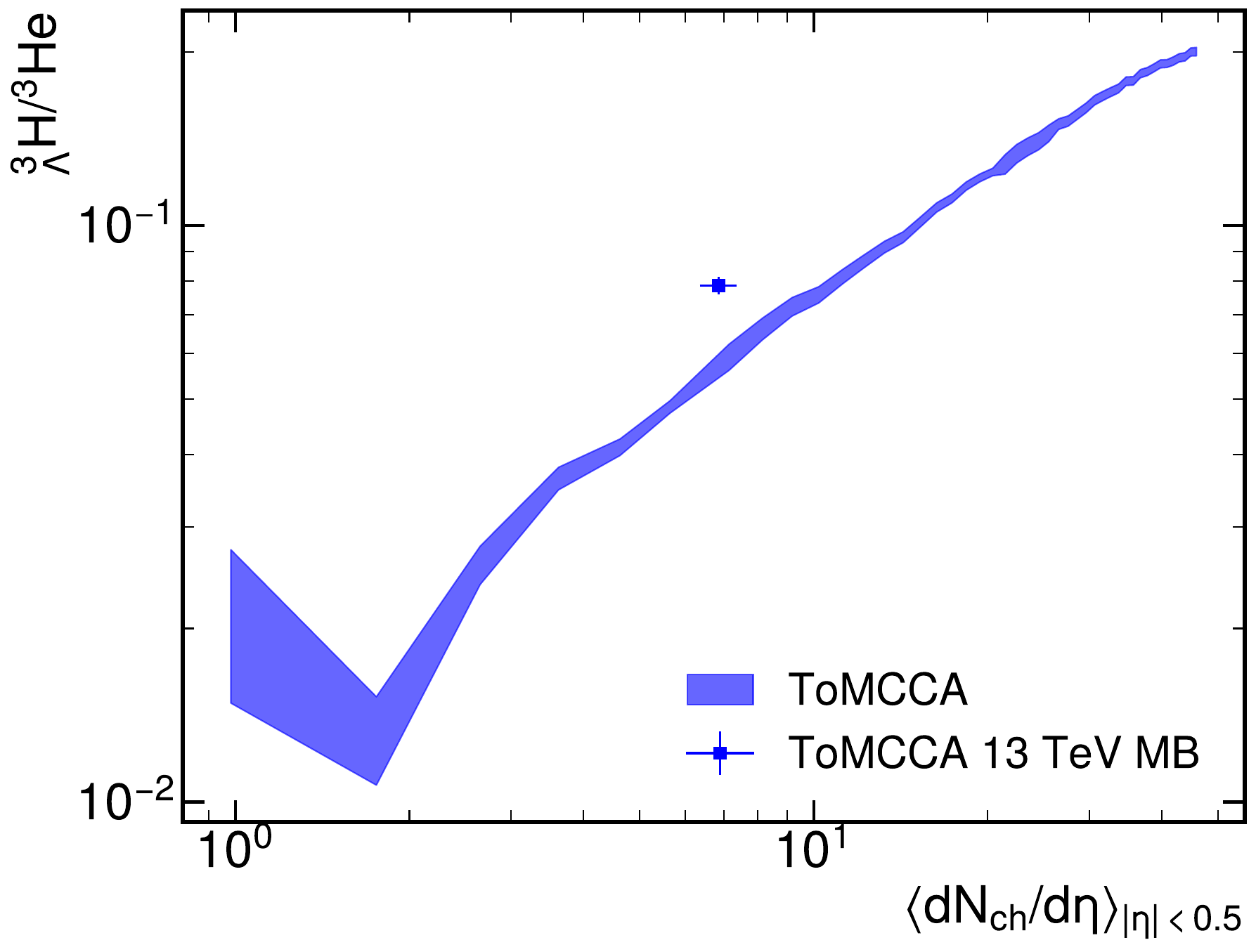}}  
    \caption{(left)$B_3$ parameter as a function of \dNetaOFive for various wave functions compared with the measurements by ALICE. In the inset, the minimum bias (INEL>0) measurements and simulations are shown. (right) \lhThree/\heThree ratio as a fuction of \dNetaOFive predicted by ToMCCA. The band indicates the combined statistical uncertainty. The square marker indicates the predictions for minimum-bias (INEL>0) collisions.}
    \label{fig:B3Mult}
\end{figure}

Fig.~\ref{fig:B3Mult} shows the measured $B_3$ parameter defined as
\begin{equation}
    B_A(p_\mathrm{T,p})=\frac{1}{2\pi p_\mathrm{T,A}}\frac{\diff^2N_A}{\diff y\diff p_\mathrm{T,A}}\bigg/\left(\frac{1}{2\pi p_\mathrm{T,p}}\frac{\diff^2N_p}{\diff y\diff p_\mathrm{T,p}}\right)^A
\end{equation}
with $A=3$ and $p_\mathrm{T,p}=p_\mathrm{T,A}/A$ for $p_\mathrm{T}/A=0.73$~GeV/$c$. In the inset, the minimum bias collisions\footnote{INEL>0 by ALICE definitions.} are enhanced for visual clarity. Overall, ToMCCA describes the measurements well, with a slight tension for minimum bias collisions which are, however covered within 2 standard deviations.
\section{Minimum bias multiplicity}
\label{sec:MinBias}
\begin{figure}[!hbt]
    \centering
    \includegraphics[width=0.99\linewidth]{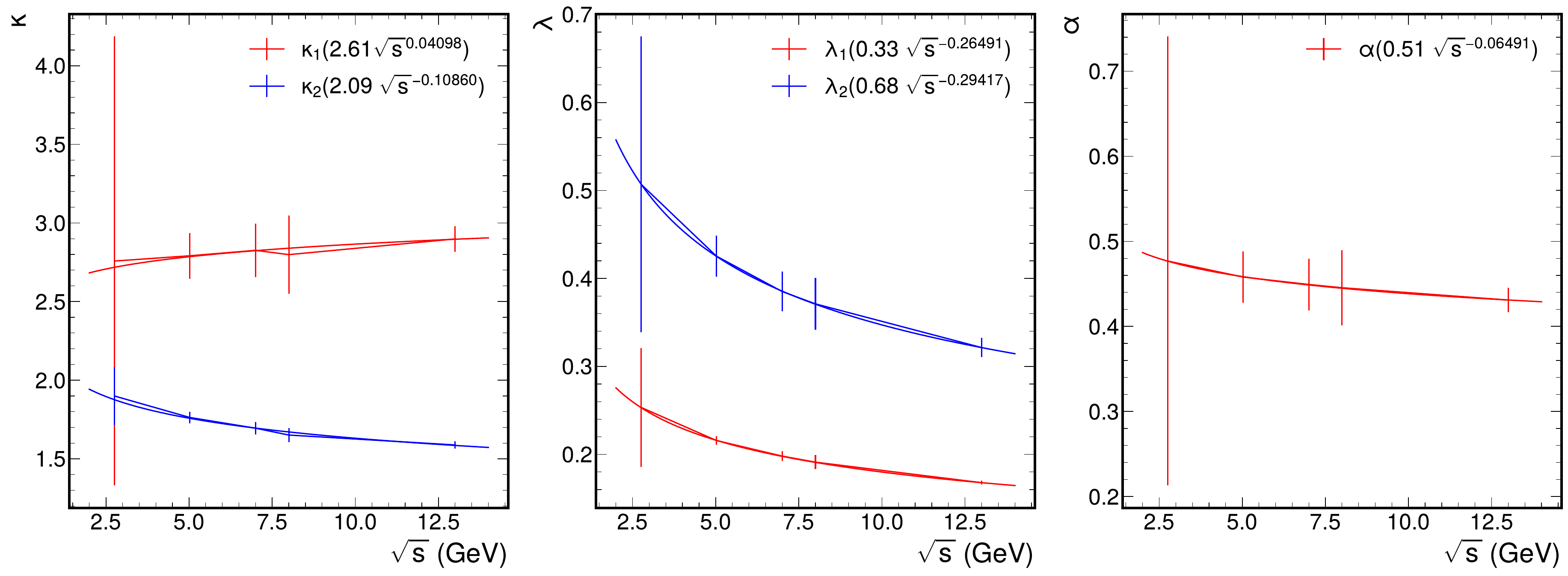}
    \caption{Parameters used in the two Erlang functions used to describe the minimum bias multiplicity distributions as a function of collision energy. }
    \label{fig:TwoErlangParams}
\end{figure}

\begin{figure}[!hbt]
    \centering
    \includegraphics[width=0.7\linewidth]{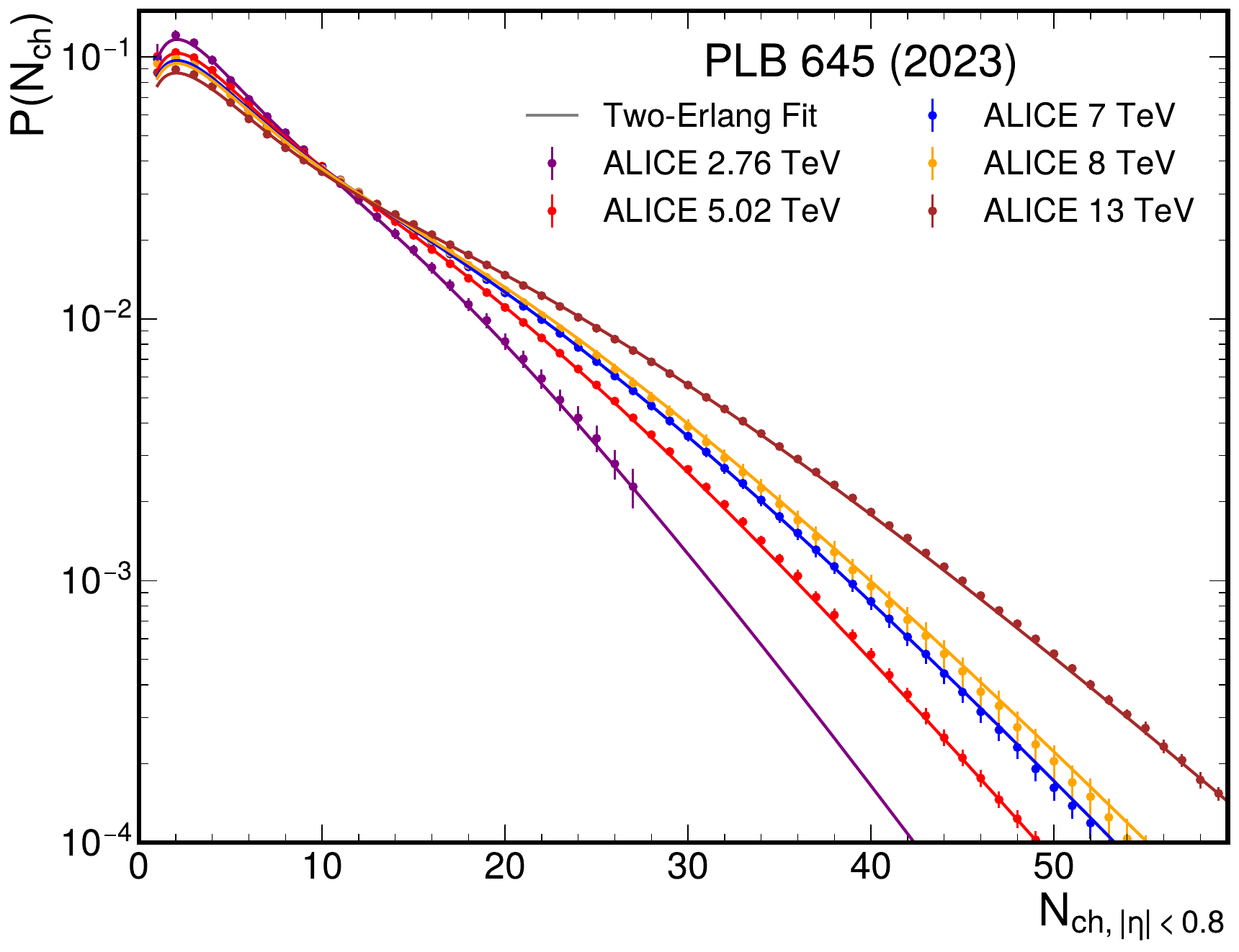}
    \caption{Published multiplicity distributions by the ALICE collaboration for $\sqrt{s}=2.76,5.02,7,8$ and $13$~TeV alongside the output of the Two-Erlang parameterization.}
    \label{fig:TwoErlangFits}
\end{figure}

Throughout this work, predictions for minimum bias collisions are made. This requires a description of the measured minimum bias distributions. For this, published distributions by the ALICE collaboration~\cite{ALICEMultiplicity} for energies from $\sqrt{s}=2.76$~TeV to $\sqrt{s}=13$~TeV are fitted using a two-Erlang distribution Ansatz
\begin{equation}
    P(N_{\rm ch}) = \alpha\frac{\lambda_1^{\kappa_1}N_{\rm ch}^{\kappa_1-1}\exp(-\lambda_1N_{\rm ch})}{\Gamma(\kappa_1)}+(1-\alpha)\frac{\lambda_2^{\kappa_2}N_{\rm ch}^{\kappa_2-1}\exp(-\lambda_2N_{\rm ch})}{\Gamma(\kappa_2)}\ .
\end{equation}
Figure~\ref{fig:TwoErlangParams} shows the obtained fit parameters as a function of the collision energy alongside parameterizations using power laws ($ax^b$). Figure~\ref{fig:TwoErlangFits} shows the measured multiplicity distributions alongside the parameterized ones.
Note that the measurements are done in a range $-0.8<\eta<0.8$ and the transformation to $\dNetaOFive$ has to be made.
\section{4D to 3D reduction}
\label{sec:4dto3dreduction}
In the so-called {\it sudden approximation}, the Lorentz-invariant yield of a trinucleon system is given by
\footnote{Note that in these calculations, we use bold and italic fonts for three- and four-vectors. For example, the four-vector $x\equiv(t,\bmr)$ and $p\equiv(p^0,\bmp)$.
Moreover, the four-vectors product will be indicate with $p\cdot r \equiv p^0 t-\bmp\cdot\bmr$.}:
\begin{multline}
\gamma \frac{\mathrm{d} N_{\mathrm{A}}}{\mathrm{~d}^3 P}=  \frac{S_{\mathrm{A}}}{(2 \pi)^4} \int \mathrm{~d}^4 x_1 \int \mathrm{~d}^4 x_2,\int \mathrm{~d}^4 x_3 \int \mathrm{~d}^4 x_1^{\prime} \int \mathrm{d}^4 x_2^{\prime}\int \mathrm{d}^4 x_3^{\prime} \\
 \times \Psi^*\left(x_1^{\prime}, x_2^{\prime}, x_3^{\prime}, \right) \Psi\left(x_1, x_2,x_3\right) \rho_{1,2,3}\left(x_1, x_2, x_3 ; x_1^{\prime}, x_2^{\prime}, x_3^{\prime}\right),
 \label{eq:firsyieldEq}
\end{multline}
where $\Psi\left(x_1, x_2,x_3\right) $ is the three-particle bound state Bethe Salpeter amplitude and $\rho_{1,2,3}$ is the three-particle reduced density matrix,
while the factor $S_\mathrm{A}$ accounts for the spin and isospin statistics of the considered case.
The density matrix for three-particle systems is assumed to be factored into single-particle densities
$\rho_{1,2,3}\left(x_1, x_2, x_3; x_1^{\prime}, x_2^{\prime}, x_3^{\prime}\right)= \rho_{1}\left(x_1;x_1^{\prime}\right)\times
\rho_{1}\left(x_2; x_2^{\prime}\right)\times \rho_{1}\left(x_3; x_3^{\prime}\right)$, and  the single-particle density is written in terms of single particle Wigner function, $f_1^W$,
\begin{eqnarray}
    \rho_1\left(x, x^{\prime}\right)=\int \frac{\mathrm{d}^4 p}{(2 \pi)^4} e^{i p\cdot\left(x^{\prime}-x\right)} f_1^W\left(p, \frac{x+x^{\prime}}{2}\right) .
\end{eqnarray}
In the following, we define the three-particle normalized Wigner density function, $W_\mathrm{A}$, as
\begin{eqnarray}
    W_\mathrm{A}\left( p_1, p_2, p_3,\frac{x_1 +x_1^\prime}{2},\frac{x_2+x_2^\prime}{2},\frac{x_3 +x_3^\prime}{2}\right) = f_1^W\left(p_1, \frac{x_1+x_1^{\prime}}{2}\right)f_1^W\left(p_2, \frac{x_2+x_2^{\prime}}{2}\right)f_1^W\left(p_3, \frac{x_3+x_3^{\prime}}{2}\right)\,,\label{eq:WA}
\end{eqnarray}
so that the three-particle reduced density matrix can be written as
\begin{multline}
  \rho_{1,2,3}\left(x_1, x_2, x_3; x_1^{\prime}, x_2^{\prime}, x_3^{\prime}\right)=
  \frac{1}{(2 \pi)^{12}} \int \mathrm{d}^4 p_1 \int \mathrm{d}^4 p_2 \int \mathrm{d}^4 p_3 \,e^{i p_1\cdot\left(x_1 -x_1^\prime\right)+i p_2\cdot\left(x_2 -x_2^\prime\right)+i p_3\cdot\left(x_3 -x_3^\prime\right)}\,\\
  \times W_\mathrm{A}\left( p_1, p_2, p_3,\frac{x_1 +x_1^\prime}{2},\frac{x_2+x_2^\prime}{2},\frac{x_3 +x_3^\prime}{2}\right)\,.
    \label{eq:threeNucleonDensity}
\end{multline}
With this definitions, Eq.~(\ref{eq:firsyieldEq}) reads
\begin{multline}
  \gamma \frac{\mathrm{d} N_{\mathrm{A}}}{\mathrm{~d}^3 P}=  \frac{S_{\mathrm{A}}}{(2 \pi)^{16}}\int \mathrm{d}^4 p_1 \int \mathrm{d}^4 p_2 \int \mathrm{d}^4 p_3
  \int \mathrm{d}^4 x_1 \int \mathrm{d}^4 x_2\,\int \mathrm{d}^4 x_3 \int \mathrm{~d}^4 x_1^{\prime} \int \mathrm{d}^4 x_2^{\prime}\int \mathrm{d}^4 x_3^{\prime} \\
  \times \Psi^*\left(x_1^{\prime}, x_2^{\prime}, x_3^{\prime}, \right) \Psi\left(x_1, x_2,x_3\right) \,
  e^{i p_1\cdot\left(x_1 -x_1^\prime\right)+i p_2\cdot\left(x_2 -x_2^\prime\right)+i p_3\cdot\left(x_3 -x_3^\prime\right)}\,\\
  \times W_\mathrm{A}\left( p_1, p_2, p_3,\frac{x_1 +x_1^\prime}{2},\frac{x_2+x_2^\prime}{2},\frac{x_3 +x_3^\prime}{2}\right)\,.
 \label{eq:yeildEq2}
\end{multline}

Now, we introduce the center-of-mass four vector $R$ and a pair of Jacobi four vectors $\xi_1$, $\xi_{2}$, related to $x_1,x_2,x_3$ by some linear relation which depends on the masses of the three particles.
In general, we have $x_i=R+w_{1,i}\xi_1+w_{2,i}\xi_2$. Moreover,  $\mathrm{d}^4x_1\, \mathrm{d}^4 x_2\, \mathrm{d}^4x_3= |J^{-1}|^4 \mathrm{d}^4R\, \mathrm{d}^4\xi_1\, \mathrm{d}^4\xi_2$.
The conjugate four momenta of $R,\xi_1,\xi_2$ are the total four momentum $P=p_1+p_2+p_3$ and
a pair of Jacobi four momenta $q_1,q_2$. We have $p_1\cdot x_1+p_2\cdot x_2 + p_3\cdot x_3 = P\cdot R + \xi_1\cdot q_1 +\xi_2\cdot q_2$.
In general, we find $p_i= c_i P + $ some linear combinations of $k_1$ and $k_2$, where $c_1+c_2+c_3=1$. 
Finally, $\mathrm{d}^4p_1 \mathrm{d}^4 p_2 \mathrm{d}^4p_3= |J|^{4} \mathrm{d}^4P\, \mathrm{d}^4q_1\, \mathrm{d}^4q_2$.

In general, the Bethe-Salpeter amplitude $\Psi_A$ can be factorized into the plane wave for center-of-mass motion with total momentum $P_\mathrm{A}$
and an internal wave function that depends only on two relative coordinates $\xi_1$, $\xi_{2}$ describing the motion of internal nucleons/$\Lambda$ relative to the center of mass.
\begin{eqnarray}
    \Psi_\mathrm{A}\left(x_1, x_2, x_3\right) = (2\pi)^{-4/2} e^{-i P_\mathrm{A}\cdot R}\varphi_\mathrm{A}(\xi_1,\xi_2)
    \label{eq:Wavefunction_A}\,.
\end{eqnarray}
At this point, Eq.~(\ref{eq:yeildEq2}) can be rewritten as
\begin{multline}
  \gamma \frac{\mathrm{d} N_{\mathrm{A}}}{\mathrm{d}^3 P}=  \frac{S_{\mathrm{A}}}{(2 \pi)^{20}} |J^{-1}|^4
  \int \mathrm{d}^4 P \int \mathrm{d}^4 q_1 \int \mathrm{d}^4 q_2 \int \mathrm{~d}^4 R \int \mathrm{d}^4 \xi_1\,\int \mathrm{d}^4 \xi_2
  \int \mathrm{d}^4 R^{\prime} \int \mathrm{d}^4 \xi_1^{\prime}\int \mathrm{d}^4 \xi_2^{\prime} \\
  \times e^{-i P_\mathrm{A}\cdot (R-R^\prime)} \varphi_\mathrm{A}^*\left(\xi_1^{\prime}, \xi_2^{\prime}\right) \varphi_\mathrm{A}\left(\xi_1, \xi_2\right)
    \,e^{i P\cdot\left( R-R^\prime\right)+i q_1\cdot\left(\xi_1 -\xi_1^\prime\right)+i q_2\cdot\left(\xi_2 -\xi_2^\prime\right)}\,\\
    \times W_\mathrm{A}\left( p_1, p_2, p_3,\frac{x_1 +x_1^\prime}{2},\frac{x_2+x_2^\prime}{2},\frac{x_3 +x_3^\prime}{2}\right)\,.
 \label{eq:yeildEq3}
\end{multline}
Changing variables to the four vectors $S=(R+R')/2$ and $T=R-R'$, we can now integrate over $T$ since $x_i+x_i^\prime$, $i=1,2,3$
depends only on $S$, obtaining $(2\pi)^4 \delta^{(4)}(P-P_\mathrm{A})$. After the integration of this $\delta$ over $d^4P$, we have
\begin{multline}
  \gamma \frac{\mathrm{d} N_{\mathrm{A}}}{\mathrm{~d}^3 P}=  \frac{S_{\mathrm{A}}}{(2 \pi)^{16}} |J^{-1}|^4
  \int \mathrm{d}^4 q_1 \int \mathrm{d}^4 q_2 \int \mathrm{~d}^4 S \int \mathrm{d}^4 \xi_1\,\int \mathrm{d}^4 \xi_2
  \int \mathrm{d}^4 \xi_1^{\prime}\int \mathrm{d}^4 \xi_2^{\prime} \\
  \times  \varphi_\mathrm{A}^*\left(\xi_1^{\prime}, \xi_2^{\prime}\right) \varphi_\mathrm{A}\left(\xi_1, \xi_2\right)
    \,e^{i q_1\cdot\left(\xi_1 -\xi_1^\prime\right)+i q_2\cdot\left(\xi_2 -\xi_2^\prime\right)}\,\\
    \times W_\mathrm{A}\left( p_1, p_2, p_3,\frac{x_1 +x_1^\prime}{2},\frac{x_2+x_2^\prime}{2},\frac{x_3 +x_3^\prime}{2}\right)\Bigr|_{P=P_\mathrm{A}}\,.
 \label{eq:yeildEq4}
\end{multline}
For the moment everything is exact. We will now approximate this relation in order to obtain the three-dimensinal relation
given in Eq.~(\ref{eq:yieldEq2}). In the following, we introduce the notation
$S=(t_S,\bmS)$, $R=(t_R,\bmR)$, $\xi_i=(\tau_i,\bm{\xi}_i)$, $i=1,2$, and $x_i=(t_i,\bmx_i)$, $i=1,\ldots,3$,
and similarly for the primed quantities. Clearly $t_S=(t_R+t_R')/2$ and $t_i=t_R+$ linear combination of $\tau_1$ and $\tau_2$. For more
details, see Ref.~\cite{Bellini2021}. The approximations we consider are:

\begin{itemize}

\item Low-energy approximation: in the trinucleon rest frame, the particles can be considered as non relativistic, so
  we can approximate the Bethe-Salpeter ampliture $\varphi_\mathrm{A}(\xi_1,\xi_2)$ with the non relativistic wave function
  $\varphi_{NR}(\bm{\xi}_1,\bm{\xi}_2)$. Moreover, in $W_\mathrm{A}$ we can approximate $p_i^0\approx c_i P^0$, disregarding the
  contribution of the internal energies $q_1^0$ and $q_2^0$. At this point, we can integrate over $q_1^0$ and $q_2^0$:
  \begin{equation}
    \int {dq_1^0\over 2\pi} {dq_2^0\over 2\pi} e^{i q_1^0(\tau_1-\tau_1')+i q_2^0(\tau_2-\tau_2')}= \delta(\tau_1-\tau_1')\delta(\tau_2-\tau_2') \
   \end{equation}

\item Equal time approximation: the three particles must exit at the same time from the
  interaction region in order to form a bound system. Note that in  $W_\mathrm{A}$, the combinations $(t_i+t_i')/2$ appear,
  which can be written as
  \begin{equation}
    {t_i+t_i'\over 2} = {t_R+t_R'\over 2} +\cdots = t_S+\cdots = \tilde t_i\ .
  \end{equation}
  In practice, the ``time'' $\tilde t_i$ are the $t_i$ where $t_R$ is replaced by $t_S$. 
  We can now assume that $ W_\mathrm{A}\sim \delta(\tilde t_1-t_0) \delta(\tilde t_2-t_0) \delta(\tilde t_3-t_0)$, where
  $t_0$ is some ``freeze-out'' time. We can rewrite this product as
  \begin{equation}
    \delta(\tilde t_1-t_0) \delta(\tilde t_2-t_0) \delta(\tilde t_3-t_0)= |J| \delta(t_S-t_0) \delta(\tau_1) \delta(\tau_2)\ .
  \end{equation}
\end{itemize}
This allows for the integration over the time components, obtaining
\begin{multline}
  \gamma \frac{\mathrm{d} N_{\mathrm{A}}}{\mathrm{~d}^3 P}=  \frac{S_{\mathrm{A}}}{(2 \pi)^{14}} |J^{-1}|^3
  \int \mathrm{d}^3 \bm{q}_1 \int \mathrm{d}^3 \bm{q}_2 \int \mathrm{d}^3 \bm{S} \int \mathrm{d}^3 \bm{\xi}_1\,\int \mathrm{d}^3 \bm{\xi}_2
  \int \mathrm{d}^3 \bm{\xi}_1^{\prime}\int \mathrm{d}^3 \bm{\xi}_2^{\prime} \\
  \times  \varphi^*_{NR}\left(\bm{\xi}_1^{\prime}, \bm{\xi}_2^{\prime}\right) \varphi_{NR}\left(\bm{\xi}_1, \bm{\xi}_2\right)
    \,e^{-i \bmq_1\cdot\left(\bm{\xi}_1 -\bm{\xi}_1^\prime\right)-i \bmq_2\cdot\left(\bm{\xi}_2 -\bm{\xi}_2^\prime\right)}\,\\
    \times W_\mathrm{A}\left((c_1P^0_d,\bmp_1),(c_2P^0_d,\bmp_2),(c_3P^0_d,\bmp_3),
    \frac{(t_0,\bmx_1 +\bmx_1^\prime)}{2},\frac{(t_0,\bmx_2+\bmx_2^\prime)}{2},\frac{(t_0,\bmx_3 +\bmx_3^\prime)}{2}\right)\Bigr|_{P=P_\mathrm{A}}\,.
 \label{eq:yeildEq5}
\end{multline}
At this point, $t_0$ can be taken arbitrarily as $t_0=0$. Reinserting the integration over $d^3S\times \delta^{(3)}(\bmR-\bmR')$, and going back to the integration over the
particle coordinates and momenta, we have
\begin{multline}
  \gamma \frac{\mathrm{d} N_{\mathrm{A}}}{\mathrm{d}^3 P}=  \frac{S_{\mathrm{A}}}{(2 \pi)^{14}}\int \mathrm{d}^3 \bm{p}_1 \int \mathrm{d}^3 \bm{p}_2 \int \mathrm{d}^3 \bm{p}_3
  \int \mathrm{d}^3 \bmx_1 \int \mathrm{d}^3 \bmx_2\,\int \mathrm{d}^3 \bmx_3 \int \mathrm{d}^3 \bmx_1^{\prime} \int \mathrm{d}^3 \bmx_2^{\prime}\int \mathrm{d}^3 \bmx_3^{\prime} \\
  \times \Psi_{NR}^*\left(\bmx_1^{\prime}, \bmx_2^{\prime}, \bmx_3^{\prime}, \right) \Psi_{NR}\left(\bmx_1, \bmx_2,\bmx_3\right)
  \,e^{-i \bmp_1\cdot\left(\bmx_1 -\bmx_1^\prime\right)-i \bmp_2\cdot\left(\bmx_2 -\bmx_2^\prime\right)-i \bmp_3\cdot\left(\bmx_3 -\bmx_3^\prime\right)}\,\\\
  \times W_\mathrm{A}\left( \bmp_1, \bmp_2, \bmp_3,\frac{\bmx_1 +\bmx_1^\prime}{2},\frac{\bmx_2+\bmx_2^\prime}{2},\frac{\bmx_3 +\bmx_3^\prime}{2}\right)\,.
  \label{eq:yeildEq7}
\end{multline}
where $\Psi_{NR}\left(\bmx_1, \bmx_2,\bmx_3\right)=(2\pi)^{-{3\over2}}\, e^{i \bmP\cdot\bmR} \varphi_{NR}\left(\bm{\xi}_1, \bm{\xi}_2\right)$.
Rescaling $W_A$ by a factor $(2\pi)^2$, the expression above coincides with Eq.~(\ref{eq:yieldEq2}).


\section{HH description of the three-particle wave functions }
\label{sec:hh}

In this work, the trinucleon bound-state wave functions are calculated using the pair-correlated HH (PHH) method
(a variation of the HH method), which has been described many times~\cite{Kievsky:2008es,10.3389/fphy.2020.00069}.
The interactions employed are the AV18 NN potential~\cite{Wiringa:1994wb}, augmented by the
UIX 3-body potential~\cite{Pudliner:1995wk}, and the Minnesota NN potential ~\cite{MinnesotaPotential}.
The latter interaction is a central potential, but it is able to reproduce the binding energy of \heThree and \hThree 
within 1\%.
The convergence of the wave functions and associated observables, as the binding energy, mass radius, etc., is
well achieved~\cite{Kievsky:2008es,10.3389/fphy.2020.00069}. Some of these results are given in Table~\ref{tab:b3}.

\begin{table}[!ht]

\begin{center}
  \begin{tabular}{l|ccc|ccc}
\hline    
 & \multicolumn{3}{c}{${}^3$H} & \multicolumn{3}{|c}{${}^3$He} \\
\hline
Model & $B$ (MeV) & $r_p$ (fm) & $r_M$ & $B$ (MeV) & $r_p$ (fm) & $r_M$ \\
\hline
AV18+UIX & 8.479 & 1.582 & 1.683 & 7.750 & 1.771 & 1.715\\
AV18 & 7.624 & -- & 1.766 & 6.925 & -- & 1.804\\
Minnesota  & 8.386 & 1.486 & 1.706& 7.711 & 1.708 &1.736\\
\hline
Expt.    & 8.482 & 1.60 & -- & 7.718 & 1.77 & --\\
\hline
  \end{tabular}
  \end{center}
  \caption{Binding energies and proton radii of ${}^3$H and ${}^3$He obtained from the employed nuclear interactions.}
\label{tab:b3}
\end{table}

\section{Congleton Formalism for Hypertriton}
\label{App:Congleton}
For the Congleton formalism, we use the Jacobi coordinate transformation given in Ref.~\cite{Congleton_1992}, and the total position space wave function is given as 
where $\Psi_{\hyp}$ and $\Psi_i$ for $i = 1,2,3$ are the wave functions for $\hyp$ and nucleons. $\Psi_{\hyp}$ is assumed to be factorized into the plane wave for center-of-mass motion and an internal wave function that depends only on two relative coordinates.
\begin{equation}
    \Psi_{\hyp}\left(\bm x_1,\bm x_2,\bm x_3\right) = \left(2\pi\right)^{-3/2} e^{i\bm P_{\hyp}\cdot \bm R}\varphi_{\hyp}(\bm \xi_1,\bm \xi_2)\ ,
    \label{eq:WavefunctionHe}
\end{equation}
where the relative Jacobi coordinates are defined in terms of the mass of two identical particles $m_1=m_2=M$ for proton and neutron and a third particle, $\Lambda$ with a different mass $m_3$. We define the mass ratio as $\kappa=\frac{m_3}{M}$ and writing single particle position and momentum cordinates $\bm x_i$, $\bm p_i$ for $i =1,2,3$
\begin{equation}
    \begin{aligned}
        \bm x_1 &= \frac{\kappa \bm R}{2 +\kappa}+\frac{2 \bm\xi_2}{2  +
        \kappa},\,\,\,\bm x_1^\prime = \frac{\kappa \bm R^\prime}{2 +\kappa}+\frac{2 \bm\xi_2^\prime}{2  +\kappa},\\
        \bm x_2 &= \frac{ \bm R}{2  +\kappa}+\frac{\bm\xi_1}{2}-\frac{\kappa\bm \xi_2}{2 +\kappa},\,\,\,\bm x_2^\prime =\frac{ \bm R^\prime}{2  +\kappa}+\frac{\bm\xi_1^\prime}{2}-\frac{\kappa\bm \xi_2^\prime}{2 +\kappa},\\
        \bm x_3 &= \frac{ \bm R}{2 +\kappa}-\frac{\bm \xi_1}{2}-\frac{\kappa\bm \xi_2}{2  +\kappa},\,\,\,\bm x_3^\prime =\frac{ \bm R^\prime}{2 +\kappa}-\frac{\bm \xi_1^\prime}{2}-\frac{\kappa\bm \xi_2^\prime}{2  +\kappa},
        \label{eq:JacobSpaceCord}
    \end{aligned}
\end{equation}
with momentum conjugates \begin{equation}
    \begin{aligned}
        \bm p_1 &= \frac{\kappa(2 +\kappa) \bm P}{2 +\kappa ^2}+\frac{(2  +\kappa)  \bm q_2}{2+\kappa^2}\,\\
          \bm p_2 &= \frac{(2  +\kappa)  \bm P}{2+\kappa^2}+\bm q_1-\frac{\kappa(2  +\kappa)\bm q_2}{2(2+\kappa^2)},\\
          \bm p_3 &=\frac{(2  +\kappa)  \bm P}{2+\kappa^2}-\bm q_1-\frac{\kappa(2  +\kappa)\bm q_2}{2(2+\kappa^2)}\ .          \label{eq:JacobMomCord}
    \end{aligned}
\end{equation}
With the Jacobian of the transformations, $|J| = -\frac{(2 +\kappa)^2}{2 + \kappa^2}$, the coordinates $\bm x_i, \bm p_i$ satisfies the following condition: $\bm p_1\cdot \bm x_1+\bm p_2\cdot \bm x_2+\bm p_3\cdot \bm x_3 =\bm \xi_1\cdot \bm q_1+\bm \xi_2\cdot \bm q_2+\bm R\cdot \bm P$. Finally defining $\bm R = \frac{\bm \zeta_1+\bm \zeta_2}{2}$ and $\bm R^\prime = \frac{\bm \zeta_2-\bm \zeta_1}{2}$ as well as $\mathrm{d}^3\bm x_1\, \mathrm{d}^3\bm x_2\,\mathrm{d}^3 \bm x_3\,\mathrm{d}^3 \bm x_1^\prime\, \mathrm{d}^3\bm x_2^\prime\,\mathrm{d}^3\bm x_3^\prime = \frac{|J^{-1}|^6}{8}\,\mathrm{d}^3\bm{\xi}_1 \,\mathrm{d}^3\bm{\xi}_2\,\mathrm{d}^3\bm{\xi}_1^\prime \,\mathrm{d}^3\bm{\xi}_2^\prime\,\mathrm{d}^3\bm{\zeta}_1 \,\mathrm{d}^3\bm{\zeta}_2$ with $\mathrm{d}^3\bm R\, \mathrm{d}^3\bm R^\prime = \frac{1}{8}\,\mathrm{d}^3\bm{\zeta}_1 \,\mathrm{d}^3\bm{\zeta}_2$,
we obtain the yield in terms of the Jacobi coordinates and the Hhpertriton wave function as

\begin{multline}
    \frac{\mathrm{d} N_{\hyp}}{\mathrm{d}^3 P} =\frac{S_{\mathrm{\hyp}}}{(2 \pi)^{12}}\left(\frac{|J^{-1}|}{2}\right)^3\,\int\mathrm{d}^3\bm{\xi}_1 \,\mathrm{d}^3\bm{\xi}_2\,\mathrm{d}^3\bm{\xi}_1^\prime \,\mathrm{d}^3\bm{\xi}_2^\prime\,\mathrm{d}^3\bm{\zeta}_1 \,\mathrm{d}^3\bm{\zeta}_2\,
    \,\mathrm{d}^3\bm q_1 \,\mathrm{d}^3\bm q_2\,\mathrm{d}^3\bm P\,\, e^{i\bm\zeta_1\cdot(\bm P- \bm P_{\hyp})} e^{-\frac{\left(2+\kappa^2\right) \zeta_2^2}{8\sigma^2 (2  +\kappa )^2}}
    \\\times \varphi_{\hyp}(\bm \xi_1 ,\bm \xi_2)\,\varphi^*_{\hyp}(\bm \xi_1^{\prime},\bm \xi_2^{\prime}) \,\,e^{i(\bm q_1 \cdot\bm \xi_1-\bm q_1 \cdot\bm \xi_1^\prime+\bm q_2\bm\cdot \bm\xi_2 -\bm q_2 \cdot\bm \xi_2^\prime )}\times (2\pi\sigma^2)^{-9/2}e^{\chi^2}  G_{\mathrm{np}\Lambda}(\bm p_1,\bm p_2,\bm p_3)|_{\bmP=\bmP_{\lhThree}}
    \label{eq:hyptriton1}
\end{multline}
with a substitution of $\chi^2$ defined as

\begin{equation}
\chi^2 = -\frac{4 \left((\xi_1+\xi_1^\prime)^2+2 (\xi_2+\xi_2^\prime)^2\right)+4\kappa  (\xi_1+\xi_1^\prime)^2+ \kappa^2\left(\xi_1^2+\bm\xi_1^\prime\cdot (2 \bm\xi_1+\bm\xi_1^\prime)+4 \xi_2^2+4 \bm\xi_2^\prime \cdot(2 \bm\xi_2+\bm\xi_2^\prime)\right)}{16 \sigma^2(2+\kappa)^2}\ .
\label{eq:defchi2}
\end{equation}
Performing pure $\delta$ integral $\int \mathrm{d}^3 \zeta_1\,e^{i( \bm P-\bm P_{\hyp})\cdot \bm \zeta_1} = (2\pi)^3\delta(\bm P-\bm P_{\hyp})$ and $\int \mathrm{d}^3 P$ and $\int \mathrm{d}^3 \zeta_{2}$ in Eq.~\eqref{eq:hyptriton1}, we obtain

\begin{multline}
    \frac{\mathrm{d} N_{\hyp}}{\mathrm{d}^3 P} =\frac{S_{\mathrm{\hyp}}}{(2 \pi)^{9}(2\pi\sigma^2)^{9/2}}\,\frac{8 \sqrt{2} \pi ^{3/2}{|J^{-1}|^3}\sigma^3(2  +\kappa )^3}{\left(2 +\kappa^2\right)^{3/2}}\,\int\mathrm{d}^3\bm{\xi}_1 \,\mathrm{d}^3\bm{\xi}_2\,\mathrm{d}^3\bm{\xi}_1^\prime \,\mathrm{d}^3\bm{\xi}_2^\prime\,\,\mathrm{d}^3\bm q_1 \,\mathrm{d}^3\bm q_2\,\varphi_{\hyp}(\bm \xi_1 ,\bm \xi_2)\,\varphi^*_{\hyp}(\bm \xi_1^{\prime},\bm \xi_2^{\prime})\\\times e^{i(\bm q_1 \cdot(\bm \xi_1-\bm \xi_1^\prime)+\bm q_2\cdot(\bm \xi_2 -\bm \xi_2^\prime))} \times e^{\chi^2}  G_{\mathrm{np}\Lambda}(\bm p_1,\bm p_2,\bm p_3)|_{\bmP=\bmP_{\lhThree}}
    \label{eq:hyptriton2}
\end{multline}
Further simplifications are made by making the following coordinate transformations $\bm\xi_1= \bm r_1 +\bm \zeta_1/2 $, $\bm \xi_1^\prime= \bm r_1 - \bm \zeta_1/2$,  and $\bm \xi_2= \bm r_2 +\bm \zeta_2/2 $, $\bm \xi_2^\prime= \bm r_2 - \bm \zeta_2/2$, with $\mathrm{d}^3\bm{\xi}_1\, \mathrm{d}^3\bm{\xi}_2\,\mathrm{d}^3\bm{\xi}_1^\prime\,\mathrm{d}^3\bm{\xi}_2^\prime = \,\mathrm{d}^3\bm{r}_1 \,\mathrm{d}^3\bm{r}_2\,\mathrm{d}^3\bm{\zeta}_1 \,\mathrm{d}^3\bm{\zeta}_2$. In this way we obtain

\begin{multline}
    \frac{\mathrm{d} N_{\hyp}}{\mathrm{d}^3 P} =\frac{S_{\mathrm{\hyp}}}{(2 \pi)^{9}(2\pi\sigma^2)^{9/2}}\,\frac{8 \sqrt{2} \pi ^{3/2}{|J^{-1}|^3}\sigma^3(2  +\kappa )^3}{\left(2 +\kappa^2\right)^{3/2}}\,\int\,\mathrm{d}^3\bm{r}_1 \,\mathrm{d}^3\bm{r}_2\,\mathrm{d}^3\bm{\zeta}_1 \,\mathrm{d}^3\bm{\zeta}_2\,\,\mathrm{d}^3\bm q_1 \,\mathrm{d}^3\bm q_2\,\,\\\times\varphi_{\hyp}(\bm r_1 +\frac{\bm\zeta_1}{2} ,\bm r_2 +\frac{\bm\zeta_2}{2})\,\varphi^*_{\hyp}(\bm r_1 -\frac{\bm \zeta_1}{2} ,\bm r_2 -\frac{\bm \zeta_2}{2}) e^{i(\bm q_1 \cdot\bm\zeta_1+\bm q_2\bm\cdot \bm\zeta_2)}e^{-\frac{r_1^2(\kappa+2)^2+4 r_2^2\left(2 +\kappa^2\right)}{4 \sigma^2(2+\kappa)^2}}  G_{\mathrm{np}\Lambda}(\bm p_1,\bm p_2,\bm p_3)|_{\bmP=\bmP_{\lhThree}}\ ,
    \label{eq:hyptriton3}
\end{multline}
where 

\begin{equation}
    \mathcal{D}(\bm q_{1},\bm q_{2},\bm r_{1},\bm r_{2})=
   \int \mathrm{d}^3 \bm{\zeta}_1 \int \mathrm{d}^3 \bm{\zeta}_2\,\, \,\varphi_{\hyp}(\bm r_1 +\frac{\bm\zeta_1}{2} ,\bm r_2 +\frac{\bm\zeta_2}{2})\,\varphi^*_{\hyp}(\bm r_1 -\frac{\bm\zeta_1}{2} ,\bm r_2 -\frac{\bm\zeta_2}{2}) e^{i(\bm q_1 \cdot\bm\zeta_1+\bm q_2\bm\cdot\bm\zeta_2)}\ .
    \label{eq:WigDensity}
\end{equation}
Within the approach developed in Ref.~\cite{Congleton_1992}, the wave function  $\varphi_{\hyp}$ is defined as the product of $\Lambda$--d wave function and deuteron wave function, assuming deuteron is undisturbed bound object at the core. Therefore 
\begin{equation}
    \varphi_{\hyp}=\varphi_{\Lambda}\varphi_d.
      \label{eq:Wavehype1}
\end{equation}
The Wigner density can be written as the product of the Wigner densities for the $\Lambda$ and  deuteron  as follows:
\begin{equation}
\begin{aligned}
    \mathcal{D}(\bm q_{1},\bm q_{2},\bm r_{1},\bm r_{2})&=
   \int \mathrm{d}^3 \bm{\zeta}_1 \int \mathrm{d}^3 \bm{\zeta}_2\,\, \,\varphi_{\Lambda}(\bm r_1 +\frac{\bm \zeta_1}{2}) \varphi^*_{\Lambda}(\bm r_1 -\frac{\bm \zeta_1}{2}),\varphi_{d}(\bm r_2 +\frac{\bm \zeta_2}{2})\varphi^*_{d}(\bm r_2 -\frac{\bm \zeta_2}{2}) e^{i(\bm q_1 \cdot\bm \zeta_1+\bm q_2\bm\cdot\bm \zeta_2)}\\
   &=\int \mathrm{d}^3 \bm{\zeta_1}\, \,\varphi_{\Lambda}(\bm r_1 +\frac{\bm \zeta_1}{2}) \varphi^*_{\Lambda}(\bm r_1 -\frac{\bm \zeta_1}{2})e^{i\bm q_1 \cdot\bm \zeta_1}\,\int \mathrm{d}^3 \bm \zeta_2\,\varphi_{d}(\bm r_2 +\frac{\bm \zeta_2}{2})\varphi^*_{d}(\bm r_2 -\frac{\bm \zeta_2}{2}) e^{i\bm q_2\bm\cdot\bm \zeta_2}\\
   &=\mathcal{D}_{\Lambda}(\bm q_{1},\bm r_{1})\mathcal{D}_d(\bm q_{2},\bm r_{2})\ ,
    \label{eq:WigDensity2}
\end{aligned}
\end{equation}
where the Wigner densities $\mathcal{D}_d(\bm k_{2},\bm r_{2})$ is computed in coordinate space. The computation of $\mathcal{D}_d(\bm k_{2},\bm r_{2})$ has already been provided in a previous work~\cite{Mahlein2023}, using the \AV wave function, while $\mathcal{D}_{\Lambda}(\bm q_{1},\bm r_{1})$ can be easily computed using the momentum-space wave function by the Congleton approach
\begin{equation}
   \varphi_{\Lambda}(q) = N(Q)\frac{e^{-\left(\frac{q}{Q}\right)^2}}{q^2+\alpha^2}\,,
    \label{eq:WFGaussNorm}
\end{equation}
with a normalization function
\begin{equation}
    N(Q) = \left\{\frac{\pi }{4 \alpha } \left[\left(\frac{4 \alpha ^2}{Q^2}+1\right) \text{Cerfe}\left(\frac{\sqrt{2} \alpha }{Q}\right)-\frac{2 \alpha  \left(\frac{2}{\pi }\right)^{1/2}}{Q}\right]\right\}^{-1/2}\ .
    \label{eq:Norm}
\end{equation}
Here $\alpha = 0.068$ fm$^{-1}$, $N^2(2.5~\mathrm{fm}^{-1})= 0.094$ and $\int |\phi_{\Lambda}(q)|^2 \diff^3\bm q = 4\pi$.
Finally, the Wigner density of the hypertriton $\Lambda$ part is given by 
\begin{equation}
   \mathcal{D}_{\Lambda}(\bm q,\bm r) = N(Q)^2\int_0^{\infty} dk \frac{4 i \pi  \left(-1+e^{2 i k r}\right)}{q r \left(k^2+4 \left(\alpha ^2+q^2\right)\right)}\log \left(\frac{4 \alpha ^2+(k-2 q)^2}{4 \alpha ^2+(k+2 q)^2}\right) e^{-\frac{k^2+2 i k Q^2 r+4 q^2}{2 Q^2}}\ ,
    \label{eq:WigLambda}
\end{equation}
and $\mathcal{D}_{\Lambda}(0,0)/4\pi = 8.0$ as expected. The probability is finally written  as 

\begin{equation}
\begin{aligned}
    \mathcal{P}( q_{1}, q_{2},\sigma)&=
    \frac{S_{\mathrm{\hyp}}}{(2\pi)^{3}\sigma^6}|J|^{-3}\,\frac{(2 +\kappa)^3}{(2 +\kappa ^2)^{3/2}}\,
   \int \mathrm{d}^3 \bm r_1 \int \mathrm{d}^3 \bm r_2\,\, \mathcal{D}_{\Lambda}(\bm q_{1},\bm r_{1})\mathcal{D}_d(\bm q_{2},\bm r_{2})e^{-\frac{r_1^2(2+\kappa)^2+4 r_2^2\left(2 +\kappa^2\right)}{4 \sigma^2(2+\kappa)^2}}\\
    \label{eq:ProbHyp}
    &= \frac{\left(2 +\kappa^2\right)^{3/2}}{(2  +\kappa)^3}\times  \frac{S_{\mathrm{\hyp}}}{(2\pi)^{3}\sigma^6}\,\,
   \int \mathrm{d}^3 \bm r_1 \int \mathrm{d}^3 \bm r_2\,\, \mathcal{D}_{\Lambda}(\bm q_{1},\bm r_{1})\mathcal{D}_d(\bm q_{2},\bm r_{2})e^{-\frac{r_1^2(2 +\kappa)^2+4 r_2^2\left(2 +\kappa^2\right)}{4 \sigma^2(2+\kappa)^2}}
    \end{aligned}
\end{equation}
with $\int \mathrm{d}^3 \bm{q}_1\,\mathrm{d}^3 \bm{q}_2\, \frac{|J|^3}{(2\pi)^9}\,  G_{\mathrm{np}\Lambda}(\bm p_1,\bm p_2,\bm p_3)|_{\bmP=\bmP_{\lhThree}} = 1$.

\bibliographystyle{utphys}
\bibliography{bibliography}

\end{document}